\pdfoutput=1
\documentclass[11pt]{article}

\usepackage[a4paper, margin=2.5cm]{geometry}
\usepackage[T1]{fontenc}
\usepackage{lmodern}
\usepackage{amsmath, amssymb, amsthm}
\usepackage{mathtools}
\usepackage{tikz}
\usepackage{enumitem}
\usepackage[authoryear,round]{natbib}
\usepackage{microtype}
\usepackage{xcolor}
\usepackage[colorlinks=true, allcolors=blue!60!black]{hyperref}

\ifdefined\AddToHook\AddToHook{para/begin}{\looseness=-1}\fi

\theoremstyle{plain}
\newtheorem{theorem}{Theorem}
\newtheorem{lemma}[theorem]{Lemma}
\newtheorem{proposition}[theorem]{Proposition}
\newtheorem{corollary}[theorem]{Corollary}
\theoremstyle{definition}
\newtheorem{definition}[theorem]{Definition}
\newtheorem{example}{Example}

\newcommand{\ci}[1]{[\![#1]\!]}
\newcommand{\st}[3]{(#1#2\,|\,#3)}
\newcommand{\apm}[1]{a_{#1}}
\newcommand{\PD}{\mathrm{PD}_n}

\newcommand{\cI}{\mathcal{I}}
\newcommand{\cM}{\mathcal{M}}
\newcommand{\R}{\mathbb{R}}
\newcommand{\indep}{\mathrel{\perp\!\!\!\perp}}
\newcommand{\edge}[2]{#1\,\text{--}\,#2}
\DeclareMathOperator{\rank}{rank}
\DeclareMathOperator{\nei}{ne}
\DeclareMathOperator{\pa}{pa}
\DeclareMathOperator{\chd}{ch}

\DeclareMathOperator{\diag}{diag}
\DeclareMathOperator{\adj}{adj}

\title{Schur multipliers, vanishing minors and\\ Gaussian conditional independence}
\author{Manuele Leonelli\\[4pt]
{\normalsize School of Science and Technology, IE University}\\
{\normalsize Paseo de la Castellana 259E, 28046 Madrid, Spain}\\
{\normalsize \href{mailto:manuele.leonelli@ie.edu}{\texttt{manuele.leonelli@ie.edu}}}}
\date{}

\begin{document}

\maketitle

\begin{abstract}
\noindent Conditional independence in a Gaussian vector is the vanishing of almost-principal minors of its covariance matrix $\Sigma$. A sensitivity analysis replaces $\Sigma$ by the Schur product $\Delta\circ\Sigma$, where the symmetric multiplier $\Delta$ differs from one exactly at the changed positions. Such a change generally destroys some conditional independences, and we determine which minors stay zero. A vanishing minor is preserved for every matrix exactly when the corresponding block of $\Delta$ has rank one. When variances or covariances change, each almost-principal minor of $\Delta\circ\Sigma$ is a polynomial in the amounts added, with minors of $\Sigma$ as coefficients, and a conditional independence survives every such change exactly when they vanish. In an undirected graphical model, changing variances joins the neighbours of the perturbed vertices, while changing covariances leaves a structure read from the graph by a separation rule, possibly not graphical. Both extend to ancestral graphs, directed acyclic graphs included, and identify the entries that can vary freely. For the other entries, blocks of rank one along a junction tree characterize the multipliers that preserve a decomposable model. A sensitivity analysis can therefore rest on the conditional independence structure after the change, as illustrated on examination marks of 88 students.


\end{abstract}

\section{Introduction}

Many statistical models for a Gaussian vector are defined by the vanishing of minors of its covariance matrix $\Sigma$, because two components are independent given a set of others exactly when an almost-principal minor of $\Sigma$ is zero \citep{drton2009}. In a graphical model a missing edge stands for a conditional independence between two variables. For an undirected graph these relations amount to a zero pattern of $\Sigma^{-1}$ \citep{lauritzen1996}, while directed acyclic graphs and the wider class of ancestral graphs describe further patterns \citep{sadeghi2014}. Since the graph is what is interpreted in applications, every relation it displays must hold for the matrix.

In practice the entries of $\Sigma$ are estimated from data or elicited from experts, so they are seldom known exactly, and a sensitivity analysis studies what happens when some of them are changed. Vanishing minors are polynomial equations, which makes them fragile, and a change of one entry of $\Sigma$ generally moves the matrix off the model. Some conditional independence relations are then lost, although the graph still displays them. It is therefore natural to ask which relations survive and whether some other graph describes the perturbed matrix. Both questions are purely matrix-theoretic once the perturbed matrix is written as a Schur product $\Delta\circ\Sigma$ with a symmetric multiplier $\Delta$, since they ask how the vanishing minors of $\Sigma$ compare with those of $\Delta\circ\Sigma$.

Sensitivity analysis for Gaussian Bayesian networks has mostly dealt with the size of the change. The conditional means and variances of such a network are rational functions of its parameters, so the effect of a change on them can be computed symbolically \citep{castillo2003}. The standard methods perturb the mean vector and the covariance matrix additively, then measure the effect with the Kullback--Leibler or a related divergence \citep{gomez2008}, which shows how far the distribution has moved without telling which relations were lost. Model-preserving sensitivity analysis addresses the structure instead, by expressing the perturbation through a Schur multiplier that also changes further entries, chosen so that the perturbed matrix stays in the model \citep{gorgen2020}. Several patterns of such changes, called covariations, are known to suffice, at the cost of modifying entries that were not meant to vary. The converse situation, in which an entry is changed with no compensation, has been addressed only for measurement error. Additive noise on a variable increases its variance, which is known to destroy the relations in which that variable is conditioned on \citep{saeed2020}. A decrease of a variance or a change of a covariance has no such interpretation.

The problem belongs to the algebraic study of Gaussian conditional independence. The minors that can vanish together are constrained by the gaussoid axioms \citep{lnenicka2007,boege2019}. Trek separation extends the description to rank conditions on arbitrary submatrices \citep{sullivant2010} and to nested determinants \citep{drton2020}. For directed acyclic graphs the hypersurface on which one such minor vanishes has been studied in connection with partial correlation testing \citep{lin2014}. These works describe the zero patterns of one matrix at a time, whereas a sensitivity analysis compares those of two. Tools for this comparison come from matrix analysis. The Schur product, also called the Hadamard product, has a long history in multivariate statistics \citep{styan1973}, and by the Schur product theorem it preserves positive semidefiniteness \citep{horn2013}. A change of one entry has rank at most two, so the Sherman--Morrison--Woodbury formula gives the new inverse \citep{horn2013}. A change of an off-diagonal entry also moves along the positive definite completions of a partial matrix, which underlie Gaussian Markov combinations \citep{massa2017}.

This paper compares the vanishing minors of $\Sigma$ and of $\Delta\circ\Sigma$ in two steps. The first asks when a multiplier keeps a minor at zero whatever matrix it is applied to. This happens exactly when the block of $\Delta$ in the rows and columns of the minor has rank one, in which case the minor of $\Delta\circ\Sigma$ is the minor of $\Sigma$ times a non-zero constant. The covariation schemes of \citet{gorgen2020} are instances of this condition. The second step considers changes of variances and covariances with no covariation. Each almost-principal minor of $\Delta\circ\Sigma$ is then a polynomial in the amounts added, with minors of $\Sigma$ as coefficients, of degree at most one in each variance and at most two in each covariance. A generic change therefore only removes relations, while new ones can appear only at the roots of these polynomials. For variances the surviving relations are determined by the original structure alone. The same holds for covariances, changed one at a time, as long as no cancellation of partial covariances occurs.

On a graphical model these rules can be read off the graph itself. A change of variances turns an undirected graphical model into another one, whose graph is obtained by joining the neighbours of the perturbed vertices, and ancestral graphs admit a similar description. After a change of covariances, the surviving relations are read off the graph by separation, with each changed covariance acting as an additional edge whose two vertices can no longer separate. This rule holds for every ancestral graph, including directed acyclic graphs, where no such cancellation can occur, and it does not depend on the order in which the covariances are changed. The resulting structure, however, need not be graphical. The graph also identifies the entries that can be varied with no compensation, namely the variances of simplicial vertices and the covariances of adjacent pairs of them, in an undirected graph or in the moral graphs of the ancestral sets of a directed acyclic graph. For the other entries of a decomposable model, the multipliers that preserve it are those with a block of rank one for each edge of a junction tree.

For matrix theory the results describe the Schur multipliers that preserve a determinantal hypersurface of symmetric matrices, and they relate the perturbation of an entry to an operation on the graph of the inverse. For statistics they allow a sensitivity analysis to rest on the conditional independence structure that holds after the change. In an application to the marks of 88 students in five mathematics examinations, a change of one standard error in a single covariance leaves only four of the sixteen relations of the fitted model. All proofs are given in Appendix~\ref{app:proofs}.

\section{Setting and definitions}\label{sec:setting}

\subsection{Conditional independence and minors}

For Gaussian vectors conditional independence is an algebraic property of the covariance matrix. We write $[n]=\{1,\dots,n\}$ and $\PD$ for the cone of symmetric positive definite $n\times n$ matrices. For $R,S\subseteq[n]$ the submatrix of $X$ with rows in $R$ and columns in $S$ is denoted by $X_{R,S}$, abbreviated to $X_R$ when $R=S$. When $|R|=|S|$ the determinant $\det X_{R,S}$ is a minor of $X$, which is principal if $R=S$ and almost-principal if the two sets differ in exactly one index \citep{fallat2018}.

Let $Y=(Y_1,\dots,Y_n)$ be a Gaussian vector with covariance matrix $\Sigma=(\sigma_{ij})\in\PD$. For disjoint $A,B,C\subseteq[n]$ we write $A\indep B\,|\,C$ when $Y_A$ and $Y_B$ are conditionally independent given $Y_C$. This holds exactly when $\Sigma_{A\cup C,B\cup C}$ has rank $|C|$ \citep{drton2009}, or equivalently when $i\indep j\,|\,C$ for all $i\in A$ and $j\in B$. We therefore work with elementary statements $\st{i}{j}{C}$, with $i\neq j$ and $C\subseteq[n]\setminus\{i,j\}$, which hold for $\Sigma$ when $i\indep j\,|\,C$. We write
\[
\apm{ij|C}(X)=\det X_{iC,jC},
\]
where $iC$ stands for $\{i\}\cup C$ with $i$ listed first and $C$ in increasing order. Every almost-principal minor is of this form up to sign. Since $\det\Sigma_C>0$, with the convention $\det\Sigma_\emptyset=1$, the submatrix $\Sigma_{iC,jC}$ has rank at least $|C|$, and the rank condition takes the form
\begin{equation}\label{eq:ci}
\st{i}{j}{C}\ \text{holds for}\ \Sigma\quad\Longleftrightarrow\quad \apm{ij|C}(\Sigma)=0.
\end{equation}
The quotient $\sigma_{ij\cdot C}=\apm{ij|C}(\Sigma)/\det\Sigma_C$ is the partial covariance of $Y_i$ and $Y_j$ given $Y_C$. The \emph{conditional independence structure} of $\Sigma$ is the set $\ci{\Sigma}$ of all elementary statements that hold for $\Sigma$, so that it records the vanishing almost-principal minors of $\Sigma$.

\begin{example}\label{ex:kite}
Let $n=4$ and consider the covariance matrix
\[
\Sigma=\begin{pmatrix}4&1&3&1\\1&4&1&1\\3&1&4&1\\1&1&1&1\end{pmatrix}.
\]
Two of its almost-principal minors are
\[
\apm{12|4}(\Sigma)=\det\begin{pmatrix}\sigma_{12}&\sigma_{14}\\ \sigma_{42}&\sigma_{44}\end{pmatrix}=\sigma_{12}\sigma_{44}-\sigma_{14}\sigma_{42}=0,\qquad \apm{23|4}(\Sigma)=\sigma_{23}\sigma_{44}-\sigma_{24}\sigma_{43}=0.
\]
Hence $Y_2$ is conditionally independent of $(Y_1,Y_3)$ given $Y_4$. Computing the $24$ almost-principal minors $\apm{ij|C}$ shows that exactly four vanish, so that $\ci{\Sigma}=\{\st{1}{2}{4},\allowbreak\ \st{1}{2}{34},\allowbreak\ \st{2}{3}{4},\allowbreak\ \st{2}{3}{14}\}$.
\end{example}

\begin{example}\label{ex:path}
Let $n=4$ and let $\Sigma$ have entries $\sigma_{ij}=2^{3-|i-j|}$, that is,
\[
\Sigma=\begin{pmatrix}8&4&2&1\\4&8&4&2\\2&4&8&4\\1&2&4&8\end{pmatrix}.
\]
Here $\apm{13|2}(\Sigma)=\sigma_{13}\sigma_{22}-\sigma_{12}\sigma_{23}=2\cdot8-4\cdot4=0$, so $Y_1$ and $Y_3$ are conditionally independent given $Y_2$. In total seven statements hold, so that the structure of this matrix is $\ci{\Sigma}=\{\st{1}{3}{2},\allowbreak\ \st{1}{3}{24},\allowbreak\ \st{2}{4}{3},\allowbreak\ \st{2}{4}{13},\allowbreak\ \st{1}{4}{2},\allowbreak\ \st{1}{4}{3},\allowbreak\ \st{1}{4}{23}\}$.
\end{example}

\subsection{Graphical models}\label{sec:graphical}

A graph displays a conditional independence structure by means of separation. %
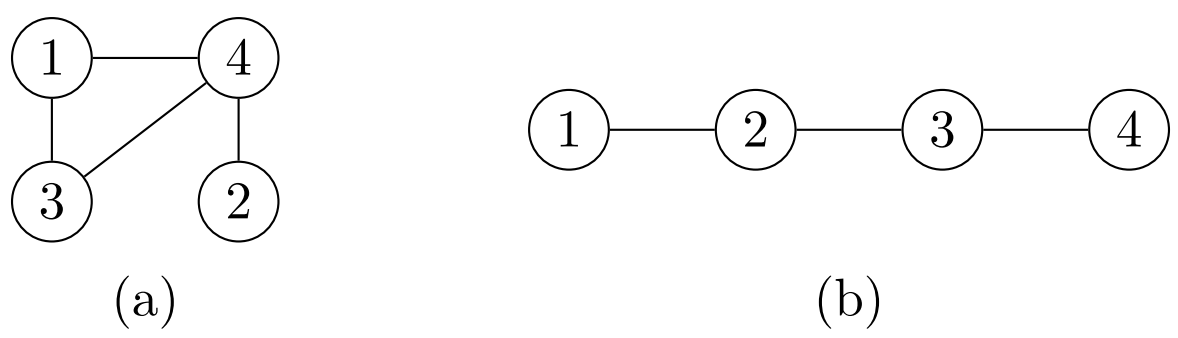
\begin{figure}[t]
\centering
\begin{tikzpicture}[every node/.style={circle,draw,inner sep=1.5pt,minimum size=15pt,font=\small},scale=0.95]
\begin{scope}
\node (1) at (0,1) {1}; \node (4) at (1.3,1) {4}; \node (3) at (0,0) {3}; \node (2) at (1.3,0) {2};
\draw (1)--(3) (1)--(4) (2)--(4) (3)--(4);
\node[draw=none] at (0.65,-0.7) {(a)};
\end{scope}
\begin{scope}[xshift=3.6cm]
\node (1) at (0,0.5) {1}; \node (2) at (1.3,0.5) {2}; \node (3) at (2.6,0.5) {3}; \node (4) at (3.9,0.5) {4};
\draw (1)--(2)--(3)--(4);
\node[draw=none] at (1.95,-0.7) {(b)};
\end{scope}
\end{tikzpicture}
\caption{The graphs of the two running examples, with (a) the graph of Example~\ref{ex:kite} and (b) the path of Example~\ref{ex:path}.}
\label{fig:graphs}
\end{figure}%
Only undirected graphs are needed until Section~\ref{sec:ancestral}, where ancestral graphs are introduced. Let $G=([n],E)$ be an undirected graph. A set $C$ separates $i$ from $j$ if every path between them contains a vertex of $C$, and $\cI(G)$ denotes the set of statements $\st{i}{j}{C}$ such that $C$ separates $i$ from $j$ in $G$. The Gaussian graphical model of $G$ is the set of positive definite matrices
\[
\cM(G)=\{\Sigma\in\PD:(\Sigma^{-1})_{ij}=0\ \text{whenever}\ i\neq j\ \text{and}\ \{i,j\}\notin E\}.
\]
The set $\cI(G)$ is defined from the graph alone. Its link with matrices is the global Markov property, by which every $\Sigma\in\cM(G)$ satisfies all the statements of $\cI(G)$ \citep{lauritzen1996}. Such a matrix is \emph{faithful} to $G$ if it satisfies no other statement, that is, if $\ci{\Sigma}=\cI(G)$. More generally, a matrix $\Sigma\in\PD$ is \emph{Markov} to a graph $G$ if it satisfies all the statements of $\cI(G)$, and the matrices that are Markov to an undirected graph are those of $\cM(G)$. Faithfulness is automatic when all off-diagonal entries of $\Sigma^{-1}$ are non-positive, in which case $\Sigma$ is faithful to the graph of the non-zero off-diagonal entries of its inverse \citep{fallat2017}. We write $\nei(k)$ for the set of neighbours of a vertex $k$ and $N[k]=\nei(k)\cup\{k\}$.

\begin{example}\label{ex:graphs}
The matrices of Examples~\ref{ex:kite} and~\ref{ex:path} are faithful to the graphs of Figure~\ref{fig:graphs}. In the first graph vertex $4$ separates $2$ from the other vertices, so the inverse has zeros exactly in positions $(1,2)$ and $(2,3)$. For the path the inverse is tridiagonal, with zeros in positions $(1,3)$, $(1,4)$ and $(2,4)$.
\end{example}

\subsection{Multipliers and perturbations}\label{sec:multipliers}

A sensitivity analysis replaces some entries of $\Sigma$ by new values. Such a change can be written as an entrywise product, which rescales several entries at once by possibly different factors, or as a sum, which adds an amount to each changed entry. We use both forms, the multiplicative one to describe which relations are preserved and the additive one to compute how the minors change.

The Schur product of two matrices of equal size is their entrywise product, $\Delta\circ X=(\delta_{ab}x_{ab})$. A \emph{multiplier} is a symmetric $n\times n$ matrix $\Delta$ with non-zero entries and positive diagonal. The diagonal must be positive because the variances of $\Delta\circ\Sigma$ have to stay positive. The entries are non-zero so that the map $\Sigma\mapsto\Delta\circ\Sigma$ is invertible, its inverse being defined by the entrywise reciprocal of $\Delta$. The map is considered on the subset of $\PD$ given by
\[
U_\Delta=\{\Sigma\in\PD:\Delta\circ\Sigma\in\PD\}
\]
of covariance matrices whose image is again a covariance matrix. If $\Delta$ is positive semidefinite, it is the whole cone $\PD$ by the Schur product theorem \citep{horn2013}. In additive form $\Delta\circ\Sigma=\Sigma+D$, where $D=(\Delta-J)\circ\Sigma$ and $J$ is the all-ones matrix. Weyl's inequality then shows that $\Sigma\in U_\Delta$ whenever the spectral radius of $D$ is smaller than the smallest eigenvalue of $\Sigma$.

The simplest multipliers change a single parameter. Writing $E_{kl}$ for the matrix with a one in position $(k,l)$ and zeros elsewhere, a variance is changed by $\Delta=J+(\delta-1)E_{kk}$, which replaces $\sigma_{kk}$ by $\delta\sigma_{kk}$. In additive form this is $\Sigma+\tau E_{kk}$ with $\tau=(\delta-1)\sigma_{kk}$. A covariance is changed by $\Delta=J+(\delta-1)(E_{kl}+E_{lk})$, or additively by $\Sigma+\tau(E_{kl}+E_{lk})$ with $\tau=(\delta-1)\sigma_{kl}$. We call the added amount $\tau$ the \emph{size} of the perturbation. The admissible sizes, those for which the perturbed matrix is positive definite, form an open interval around zero because $\PD$ is open and convex. An increase of a variance is always admissible, since $J+(\delta-1)E_{kk}$ is positive semidefinite for $\delta>1$.

Let $\cM\subseteq\PD$ be a set of covariance matrices, for instance a graphical model or the set of matrices that satisfy a given statement. A multiplier $\Delta$ is called \emph{model-preserving} for $\cM$ if $\Delta\circ\Sigma\in\cM$ for every $\Sigma\in\cM\cap U_\Delta$. A multiplier that changes one parameter is rarely model-preserving. The approach of \citet{gorgen2020} is to add a \emph{covariation}, that is, to change further entries at the same time so that the multiplier becomes model-preserving.

\begin{example}\label{ex:covariation}
In Example~\ref{ex:kite} suppose that $\sigma_{12}$ is replaced by $\delta\sigma_{12}$ with $\delta\neq1$. Then
\[
\apm{12|4}(\Delta\circ\Sigma)=\delta\sigma_{12}\sigma_{44}-\sigma_{14}\sigma_{42}=\delta-1\neq0 .
\]
The variables $Y_1$ and $Y_2$ are no longer conditionally independent given $Y_4$, so the perturbed matrix has left the model of the graph. Suppose instead that $\sigma_{23}$ and $\sigma_{24}$ are multiplied by $\delta$ as well, which corresponds to the choice of multiplier
\[
\Delta=\begin{pmatrix}1&\delta&1&1\\ \delta&1&\delta&\delta\\ 1&\delta&1&1\\ 1&\delta&1&1\end{pmatrix}.
\]
Now $\apm{12|4}(\Delta\circ\Sigma)=\delta\sigma_{12}\sigma_{44}-\sigma_{14}\,\delta\sigma_{42}=0$ and $\apm{23|4}(\Delta\circ\Sigma)=\delta\sigma_{23}\sigma_{44}-\delta\sigma_{24}\sigma_{43}=0$, so the model is preserved. The price is that two entries have changed that were not meant to.
\end{example}

\section{Vanishing minors under perturbation}\label{sec:main}

\subsection{Multipliers that preserve a statement}\label{sec:preserve}

Multiplying the rows and columns of a matrix by non-zero constants does not change which of its minors vanish. A multiplier acts in this way on a block $X_{R,S}$ when $\Delta_{R,S}=uv^\top$ has rank one, since then $(\Delta\circ X)_{R,S}=\diag(u)\,X_{R,S}\,\diag(v)$. The next theorem shows that this is the only mechanism.

\begin{theorem}\label{thm:rankone}
Let $\Delta$ be a multiplier and $\st{i}{j}{C}$ a statement with $C\neq\emptyset$.
\begin{enumerate}[label=\textup{(\alph*)},leftmargin=2.4em]
\item If $\rank\Delta_{iC,jC}=1$, then $\Delta\circ\Sigma$ satisfies the statement if and only if $\Sigma$ does, for every $\Sigma\in U_\Delta$.
\item If $\rank\Delta_{iC,jC}>1$, then $\Delta\circ\Sigma$ violates it for a dense open set of the $\Sigma\in U_\Delta$ that satisfy it.
\end{enumerate}
Consequently, for disjoint non-empty $A,B,C\subseteq[n]$, the multiplier $\Delta$ is model-preserving for the set of matrices that satisfy $A\indep B\,|\,C$ if and only if the block $\Delta_{A\cup C,B\cup C}$ has rank one.
\end{theorem}

A multiplier preserves a statement only when it rescales the rows and the columns of the corresponding submatrix. A statement with $C=\emptyset$ says that $\sigma_{ij}=0$ and is preserved by every multiplier. Since blocks of rank one are closed under Schur products and entrywise inverses, the multipliers that preserve a given statement form a group. Covariations chosen for different parameters can thus be applied one after the other, which is how \citet{gorgen2020} vary several parameters at the same time.

\begin{example}\label{ex:rankone}
The model of the graph in Example~\ref{ex:kite} is the set of matrices that satisfy $2\indep\{1,3\}\,|\,4$. The corresponding block of a multiplier has rows $\{2,4\}$ and columns $\{1,3,4\}$. If only $\sigma_{12}$ is multiplied by a factor $\delta$, this block takes the form
\[
\Delta_{24,134}=\begin{pmatrix}\delta&1&1\\1&1&1\end{pmatrix},
\]
which has rank two. By Theorem~\ref{thm:rankone} the multiplier is not model-preserving, in agreement with Example~\ref{ex:covariation}. The block has rank one for each of the choices
\[
\begin{pmatrix}\delta&\delta&\delta\\1&1&1\end{pmatrix},\qquad
\begin{pmatrix}\delta&1&1\\ \delta&1&1\end{pmatrix},\qquad
\begin{pmatrix}\delta&1&1\\1&\delta^{-1}&\delta^{-1}\end{pmatrix}.
\]
The first is the covariation of Example~\ref{ex:covariation}. The second multiplies $\sigma_{14}$ by $\delta$ instead. Both fill a whole row or column of the block with $\delta$, the pattern of the covariations in \citet{gorgen2020}. The third, with $\delta>0$, divides $\sigma_{34}$ and $\sigma_{44}$ by $\delta$, a covariation outside these schemes allowed by Theorem~\ref{thm:rankone}. With factors $\delta$,~$\delta'$ the Schur product of the first two has rows $(\delta\delta',\delta,\delta)$ and $(\delta',1,1)$, again of rank one.
\end{example}

\subsection{Change of variances}\label{sec:var}

From now on we write a change in the additive form of Section~\ref{sec:multipliers}, which adds an amount to each changed entry and also covers a zero covariance. These amounts are the sizes of the change, forming a vector $\tau$. Every minor of the perturbed matrix is a polynomial in $\tau$, equal to the minor of $\Sigma$ at $\tau=0$.

\begin{definition}\label{def:generic}
A property of the perturbed matrix holds for \emph{generic} $\tau$ if the sizes at which it fails lie in a proper algebraic subset of the space of sizes. This subset may depend on $\Sigma$.
\end{definition}

Such a subset is closed and has measure zero. It is finite when a single parameter changes, because a non-zero polynomial in one variable has finitely many roots. If a statement fails for $\Sigma$, its minor in the perturbed matrix is a polynomial in $\tau$ that is non-zero at $\tau=0$, so the statement fails for generic $\tau$. It can still hold at exceptional sizes, as at $\tau=-1$ in Example~\ref{ex:kitevar}. Since there are finitely many statements and a finite union of proper algebraic subsets is again proper, a generic change creates no new statement. The question is then which statements of $\Sigma$ it removes. For variances the answer comes from expanding each minor along the perturbed diagonal entries.

\begin{proposition}\label{thm:var}
Let $\Sigma\in\PD$, $K\subseteq[n]$, $\tau\in\R^K$ and $\Sigma'=\Sigma+\sum_{k\in K}\tau_kE_{kk}$. For each statement $\st{i}{j}{C}$,
\begin{equation}\label{eq:var}
\apm{ij|C}(\Sigma')=\sum_{T\subseteq K\cap C}\ \prod_{k\in T}\tau_k\ \apm{ij|C\setminus T}(\Sigma).
\end{equation}
Consequently, if $\Sigma'\in\PD$ it satisfies each statement of the set below, which equals $\ci{\Sigma'}$ for generic $\tau$,
\begin{equation}\label{eq:varstruct}
\bigl\{\st{i}{j}{C}\in\ci{\Sigma}:\ \st{i}{j}{C\setminus T}\in\ci{\Sigma}\ \text{for all}\ T\subseteq K\cap C\bigr\}.
\end{equation}
\end{proposition}

A statement survives a change of the variance of $Y_k$ when $Y_k$ is not one of its conditioning variables, or when the statement holds also without conditioning on $Y_k$. For $\tau_k>0$ the matrix $\Sigma'$ is the covariance of the vector with $Y_k$ replaced by $Y_k+\varepsilon_k$, where $\varepsilon_k$ is independent noise of variance $\tau_k$. Conditioning on this noisy version does not remove the dependence explained by $Y_k$ itself \citep{saeed2020}. By identity \eqref{eq:var}, a decrease of the variance has the same effect on the statements that condition on $Y_k$.

\begin{example}\label{ex:kitevar}
In Example~\ref{ex:kite} add $\tau$ to $\sigma_{44}$. All four statements condition on $Y_4$. Their minors vanish for $\Sigma$, so identity \eqref{eq:var} gives the four values
\[
\apm{12|4}(\Sigma')=\tau\sigma_{12},\quad \apm{12|34}(\Sigma')=\tau\,\apm{12|3}(\Sigma),\quad \apm{23|4}(\Sigma')=\tau\sigma_{23},\quad \apm{23|14}(\Sigma')=\tau\,\apm{23|1}(\Sigma),
\]
which are all equal to $\tau$. No statement survives, whatever the size. A change of $\sigma_{11}$ behaves differently. The only statement that conditions on $Y_1$ is $\st{2}{3}{14}$, and $\apm{23|14}(\Sigma')=\apm{23|14}(\Sigma)+\tau\,\apm{23|4}(\Sigma)=0$. The structure is unchanged except at $\tau=-1$, where $\st{2}{3}{1}$, $\st{3}{4}{1}$ and $\st{3}{4}{12}$ hold in addition.
\end{example}

\begin{example}\label{ex:pathvar}
In Example~\ref{ex:path} add $\tau$ to $\sigma_{22}$. Three statements are lost, since
\[
\apm{13|2}(\Sigma')=\tau\sigma_{13}=2\tau,\qquad \apm{14|2}(\Sigma')=\tau\sigma_{14}=\tau,
\]
and $\apm{13|24}(\Sigma')=\tau\,\apm{13|4}(\Sigma)=12\tau$.
The statement $\st{1}{4}{23}$ survives because $\st{1}{4}{3}$ holds, and the three statements that do not condition on $Y_2$ are untouched. The perturbed structure is $\{\st{1}{4}{3},\st{1}{4}{23},\st{2}{4}{3},\st{2}{4}{13}\}$.
\end{example}

\subsection{Change of one covariance}\label{sec:cov}

When $\sigma_{kl}$ changes, an almost-principal minor is affected only if its submatrix contains the position $(k,l)$ or $(l,k)$. This requires both $k$ and $l$ to occur in the statement, either as endpoints or as conditioning variables, which leaves three cases to consider, depending on the roles of $k$ and $l$.

\begin{proposition}\label{thm:cov}
Let $\Sigma\in\PD$, $k\neq l$ and $\Sigma'=\Sigma+\tau(E_{kl}+E_{lk})$. For every statement $s=\st{i}{j}{C}$,
\begin{equation}\label{eq:cov}
\apm{s}(\Sigma')=\apm{s}(\Sigma)+\tau\,L_s(\Sigma)-\tau^2\,Q_s(\Sigma),
\end{equation}
where $L_s=Q_s=0$ if $k$ and $l$ do not both occur in $s$. Otherwise let $D=C\setminus\{k,l\}$. Up to exchanging $k$ with $l$ or $i$ with $j$, there are three cases.
\begin{enumerate}[label=\textup{(\alph*)},leftmargin=2.4em]
\item If $\{i,j\}=\{k,l\}$ then $L_s=\det\Sigma_C$ and $Q_s=0$.
\item If $i=k$ and $l\in C$ then $L_s=-\apm{lj|D}$ and $Q_s=0$.
\item If $k,l\in C$ then $Q_s=\apm{ij|D}$ and
\begin{equation}\label{eq:L}
L_s=\det\Sigma_D\bigl(\sigma_{ik\cdot D}\,\sigma_{jl\cdot D}+\sigma_{il\cdot D}\,\sigma_{jk\cdot D}-2\,\sigma_{ij\cdot D}\,\sigma_{kl\cdot D}\bigr).
\end{equation}
\end{enumerate}
Consequently, for generic $\tau$ with $\Sigma'\in\PD$,
\begin{equation}\label{eq:covstruct}
\ci{\Sigma'}=\bigl\{s\in\ci{\Sigma}:\ L_s(\Sigma)=Q_s(\Sigma)=0\bigr\}.
\end{equation}
\end{proposition}

We refer to the three cases as types (a), (b) and (c). A statement of type (a) concerns the two variables whose covariance is changed. It never survives, because $\det\Sigma_C>0$. In type (b) one endpoint is $k$ while $l$ is a conditioning variable. The statement survives when $\st{l}{j}{D}$ holds too, that is, when $Y_j$ is also independent of $Y_l$ given the rest of $Y_C$. In type (c) both $k$ and $l$ are conditioning variables. The statement survives when it holds without them, so that $\sigma_{ij\cdot D}=0$, and when the partial covariances satisfy
\begin{equation}\label{eq:cancel}
\sigma_{ik\cdot D}\,\sigma_{jl\cdot D}+\sigma_{il\cdot D}\,\sigma_{jk\cdot D}=0 .
\end{equation}
Types (a) and (b) are thus decided by the structure of $\Sigma$, as for variances. Type (c) is decided by it only if \eqref{eq:cancel} can hold just with a zero factor in each product. Otherwise the two products may cancel, and survival depends on the values in $\Sigma$. The next definition excludes this.

\begin{definition}\label{def:L}
A set of statements $\mathcal{J}$ is \emph{cancellation-free} for a pair $\{k,l\}$ with $k\neq l$ if, for all distinct $i,j\notin\{k,l\}$ and all $D\subseteq[n]\setminus\{i,j,k,l\}$ with $\st{i}{j}{D}\in\mathcal{J}$ and $\st{i}{j}{D\cup\{k,l\}}\in\mathcal{J}$, at least one of $\st{i}{k}{D}$, $\st{i}{l}{D}$, $\st{j}{k}{D}$, $\st{j}{l}{D}$ belongs to $\mathcal{J}$.
\end{definition}

Section~\ref{sec:graphs} shows that the structure of every ancestral graph is cancellation-free. The statements that survive the change are then of two kinds.

\begin{definition}\label{def:rule}
Let $\mathcal{J}$ be a set of statements, $k\neq l$ and $\st{i}{j}{C}\in\mathcal{J}$, with $D=C\setminus\{k,l\}$. The statement is \emph{untouched} by $\{k,l\}$ if $k$ and $l$ do not both occur in it. It is \emph{screened} by an endpoint $e\in\{i,j\}\setminus\{k,l\}$ if $\st{e}{a}{D}\in\mathcal{J}$ for every $a\in\{i,j,k,l\}\setminus\{e\}$. The set $\mathcal{J}^{kl}$ consists of the statements of $\mathcal{J}$ that are untouched by $\{k,l\}$ or screened by some endpoint.
\end{definition}

The submatrix of an untouched statement does not contain $\sigma_{kl}$. A screened statement says that $Y_e$ is independent of the other variables of $\{i,j,k,l\}$ given $Y_D$. This is decided by a submatrix that does not contain $\sigma_{kl}$ either, so it survives the change, and it implies $\st{i}{j}{C}$.

\begin{proposition}\label{thm:rule}
Let $\Sigma$, $k$, $l$ and $\Sigma'$ be as in Proposition~\ref{thm:cov}, and let $\ci{\Sigma}^{kl}$ be the set of untouched and screened statements of $\ci{\Sigma}$. Every statement of $\ci{\Sigma}^{kl}$ holds for $\Sigma'$ whenever $\Sigma'\in\PD$. If $\ci{\Sigma}$ is cancellation-free for $\{k,l\}$, then $\ci{\Sigma'}=\ci{\Sigma}^{kl}$ for generic $\tau$ with $\Sigma'\in\PD$.
\end{proposition}

\begin{example}\label{ex:kitecov}
In Example~\ref{ex:kite} add $\tau$ to $\sigma_{12}$. The statement $\st{2}{3}{4}$ does not involve both $Y_1$ and $Y_2$, so it is untouched. The statements $\st{1}{2}{4}$ and $\st{1}{2}{34}$ are of type (a), with
\[
\apm{12|4}(\Sigma')=\tau\sigma_{44}=\tau,\qquad \apm{12|34}(\Sigma')=\tau\det\Sigma_{34}=3\tau .
\]
The statement $\st{2}{3}{14}$ is of type (b) with $k=2$ and $l=1$. Here $\apm{23|14}(\Sigma')=-\tau\,\apm{13|4}(\Sigma)=-2\tau$. Hence $\ci{\Sigma'}=\{\st{2}{3}{4}\}$ for generic admissible $\tau$.
\end{example}

\begin{example}\label{ex:pathcov}
In Example~\ref{ex:path} add $\tau$ to $\sigma_{23}$. The statements $\st{1}{4}{2}$ and $\st{1}{4}{3}$ are untouched, because each of them leaves out $Y_2$ or $Y_3$. Four of the other five are of type (b) and are lost. For instance $\apm{13|2}(\Sigma')=-\tau\sigma_{12}=-4\tau$. The statement $\st{1}{4}{23}$ is of type (c) with $D=\emptyset$. By \eqref{eq:L},
\[
\apm{14|23}(\Sigma')=\tau\,(\sigma_{12}\sigma_{43}+\sigma_{13}\sigma_{42}-2\sigma_{14}\sigma_{23})-\tau^2\sigma_{14}=12\tau-\tau^2 .
\]
The admissible sizes are those in the interval $(-8,24/7)$, so the root $\tau=12$ is excluded and the statement is lost. Hence $\ci{\Sigma'}=\{\st{1}{4}{2},\st{1}{4}{3}\}$ for generic admissible $\tau$. Conditioning on either interior variable makes the two ends independent, while conditioning on both of them does not.
\end{example}

Several covariances can be changed one at a time, since Propositions~\ref{thm:cov} and~\ref{thm:rule} hold for every positive definite matrix. If $\ci{\Sigma}$ and each set produced along the way are cancellation-free for the next pair, applying the rule once for each pair gives the final structure for generic sizes.

\section{Graphical models}\label{sec:graphs}

This section reads the results of Section~\ref{sec:main} on a graph to which $\Sigma$ is Markov or faithful.

\subsection{Undirected graphs}\label{sec:undirected}

For variances the perturbed matrix is again faithful to an undirected graph, obtained by adding edges. For $K\subseteq[n]$ let $G^K$ be the graph on $[n]$ in which two vertices are adjacent whenever they are joined in $G$ by a path whose interior vertices all lie in $K$. It contains $G$. For a single vertex $k$ it is obtained by joining every pair of neighbours of $k$, which is the step performed when $k$ is eliminated, except that $k$ is kept. In general $G^K$ is the elimination graph of the vertices of $K$, again with these vertices kept, and it does not depend on the order of elimination.

\begin{theorem}\label{thm:ugvar}
Let $G$ be an undirected graph, $K\subseteq[n]$ and $\Sigma'=\Sigma+\sum_{k\in K}\tau_kE_{kk}\in\PD$. If $\Sigma$ is Markov to $G$ then $\Sigma'$ is Markov to $G^K$, while if $\Sigma$ is faithful to $G$ then $\Sigma'$ is faithful to $G^K$ for generic $\tau$.
\end{theorem}

\begin{figure}[t]
\centering
\begin{tikzpicture}[every node/.style={circle,draw,inner sep=1.5pt,minimum size=15pt,font=\small},scale=0.95]
\begin{scope}
\node (1) at (0,1) {1}; \node (4) at (1.3,1) {4}; \node (3) at (0,0) {3}; \node (2) at (1.3,0) {2};
\draw (1)--(3) (1)--(4) (2)--(4) (3)--(4) (1)--(2) (2)--(3);
\node[draw=none] at (0.65,-0.7) {(a)};
\end{scope}
\begin{scope}[xshift=3.6cm]
\node (1) at (0,0.3) {1}; \node (2) at (1.3,0.3) {2}; \node (3) at (2.6,0.3) {3}; \node (4) at (3.9,0.3) {4};
\draw (1)--(2)--(3)--(4);
\draw (1) to[bend left=45] (3);
\node[draw=none] at (1.95,-0.7) {(b)};
\end{scope}
\end{tikzpicture}
\caption{Graphs after a change of one variance, (a) $G^{\{4\}}$ for Example~\ref{ex:kite} and (b) $G^{\{2\}}$ for Example~\ref{ex:path}.}
\label{fig:var}
\end{figure}
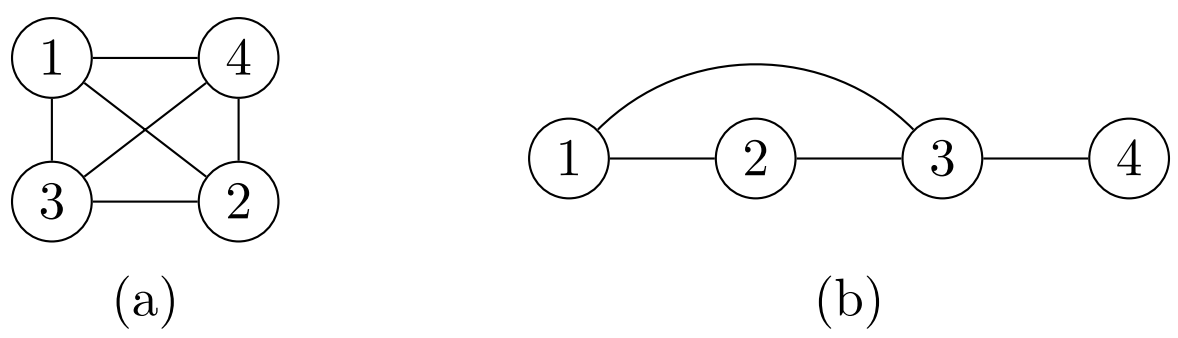

\begin{example}\label{ex:ugvar}
In Example~\ref{ex:kite} vertex $4$ is adjacent to all the others. The graph $G^{\{4\}}$ is complete and no statement survives a change of $\sigma_{44}$, as found in Example~\ref{ex:kitevar}. The neighbours of vertex $1$ are adjacent, so $G^{\{1\}}=G$ and a change of $\sigma_{11}$ has no effect. In Example~\ref{ex:path} a change of $\sigma_{22}$ joins the vertices $1$ and $3$ (Figure~\ref{fig:var}). In the new graph vertex $3$ still separates $4$ from $1$ and from $2$, which gives the four statements of Example~\ref{ex:pathvar} without any computation.
\end{example}

For one covariance the perturbed structure is described by separation in $G$ itself. Given $k\neq l$, the set $\cI(G)^{kl}$ of Definition~\ref{def:rule} consists of the untouched statements of $\cI(G)$ and of the screened ones, in which $D=C\setminus\{k,l\}$ separates an endpoint $e\notin\{k,l\}$ from the other vertices of $\{i,j,k,l\}$. Let $\cI_0$ be the set of untouched statements of $\cI(G)$, and let $G^{kl}$ be obtained from $G$ by joining every pair of vertices of $N[k]\cup N[l]$, so that this set becomes a clique. Here $G^{kl}=G^{\{k,l\}}$ exactly when $k$ and $l$ are adjacent.

\begin{theorem}\label{thm:ugcov}
Let $G$ be an undirected graph, $k\neq l$ and $\Sigma'=\Sigma+\tau(E_{kl}+E_{lk})\in\PD$. If $\Sigma$ is Markov to $G$ then $\cI(G)^{kl}\subseteq\ci{\Sigma'}$. If $\Sigma$ is faithful to $G$, the following two properties hold for generic $\tau$.
\begin{enumerate}[label=\textup{(\alph*)},leftmargin=2.4em]
\item $\ci{\Sigma'}=\cI(G)^{kl}$.
\item The graph of the non-zero off-diagonal entries of $(\Sigma')^{-1}$ is $G^{kl}$.
\end{enumerate}
Moreover, $\cI(G)^{kl}$ is the smallest of the structures $\ci{\Sigma}$ with $\Sigma\in\PD$ that contain $\cI_0$.
\end{theorem}

Unlike a change of variances, a change of one covariance can lead to a structure that no undirected graph displays (Example~\ref{ex:ugcov}). Theorem~\ref{thm:ugcov} therefore describes it as a set of statements. The statements of $\cI_0$ are untouched by the perturbation. They concern the vectors $(Y_a)_{a\neq k}$ and $(Y_a)_{a\neq l}$, whose covariance matrices do not contain $\sigma_{kl}$. The last part of the theorem says that nothing survives beyond what these two margins force. By part (b) the graph of the inverse gains all the edges among $k$, $l$ and their neighbours. The statements displayed by $G^{kl}$ hold for $\Sigma'$, but $\cI(G)^{kl}$ can be strictly larger than $\cI(G^{kl})$.

\begin{example}\label{ex:ugcov}
In Example~\ref{ex:kite} take $k=1$ and $l=2$. The statement $\st{2}{3}{4}$ is untouched. The statements $\st{1}{2}{4}$ and $\st{1}{2}{34}$ have no endpoint outside $\{1,2\}$. The statement $\st{2}{3}{14}$ is not screened, because the endpoint $3$ is adjacent to $1$. Hence $\cI(G)^{12}=\{\st{2}{3}{4}\}$, as in Example~\ref{ex:kitecov}. Since $N[1]\cup N[2]$ contains all vertices, $G^{12}$ is complete. No undirected graph displays the structure $\{\st{2}{3}{4}\}$, because in such a graph separation by $\{4\}$ always implies separation by $\{1,4\}$.

In Example~\ref{ex:path} take $k=2$ and $l=3$. Then $\cI_0=\{\st{1}{4}{2},\st{1}{4}{3}\}$. None of the other five statements is screened, since vertex $1$ is adjacent to $2$ and vertex $4$ to $3$. Hence $\cI(G)^{23}=\cI_0$, as in Example~\ref{ex:pathcov}, while $G^{23}$ is complete. The two statements come from the margins $(Y_1,Y_3,Y_4)$ and $(Y_1,Y_2,Y_4)$.
\end{example}

The next corollary uses Theorems~\ref{thm:ugvar} and~\ref{thm:ugcov} to identify the parameters that can be changed with no covariation. Recall that a vertex is simplicial if its neighbours are pairwise adjacent.

\begin{corollary}\label{cor:free}
Let $\Sigma$ be faithful to an undirected graph $G$ and let $\tau$ be a generic admissible size. The matrix $\Sigma+\tau E_{kk}$ is faithful to $G$ if and only if $k$ is simplicial. The matrix $\Sigma+\tau(E_{kl}+E_{lk})$ is faithful to $G$ if and only if $k$ and $l$ are adjacent simplicial vertices of $G$.
\end{corollary}

\begin{example}\label{ex:free}
In Example~\ref{ex:kite} the vertices $1$, $2$ and $3$ are simplicial. Among them only $1$ and $3$ are adjacent. The parameters $\sigma_{11}$, $\sigma_{22}$, $\sigma_{33}$ and $\sigma_{13}$ can therefore be varied freely, whereas a change of any other parameter alters the structure. In Example~\ref{ex:path} only the end vertices are simplicial and they are not adjacent. Only the variances $\sigma_{11}$ and $\sigma_{44}$ can be varied freely, while no covariance can.
\end{example}

Theorems~\ref{thm:ugvar} and~\ref{thm:ugcov} give the graph that a change produces, and Corollary~\ref{cor:free} identifies the parameters whose change leaves the graph as it is. For decomposable graphs, the junction tree also identifies the multipliers that preserve the model. An undirected graph $G$ is decomposable if its maximal cliques, the maximal sets of pairwise adjacent vertices, can be arranged as the nodes of a tree $T$ in which the intersection of any two of them is contained in every maximal clique on the path between them. Such a tree is called a junction tree \citep{lauritzen1996}. Removing an edge $e=\{C_1,C_2\}$ of $T$ splits it into two subtrees. Let $S_e=C_1\cap C_2$. Let $A_e$ be the set of vertices outside $S_e$ that lie in a maximal clique of the first subtree, and define $B_e$ in the same way from the second subtree. Then $S_e$ separates $A_e$ from $B_e$ in $G$, and it is non-empty when $G$ is connected.

\begin{theorem}\label{thm:decomposable}
Let $G$ be a connected decomposable graph with junction tree $T$. A multiplier $\Delta$ is model-preserving for $\cM(G)$ if and only if the block $\Delta_{A_e\cup S_e,B_e\cup S_e}$ has rank one for every edge $e$ of $T$.
\end{theorem}

Sufficiency follows from Theorem~\ref{thm:rankone}, because $\cM(G)$ is the set of matrices that satisfy $A_e\indep B_e\,|\,S_e$ for every edge $e$. Necessity does not, since the model is smaller than each of these sets.

\begin{example}\label{ex:pathmult}
The graph of Example~\ref{ex:kite} has the maximal cliques $\{1,3,4\}$ and $\{2,4\}$. Its junction tree has a single edge, so Theorem~\ref{thm:decomposable} reduces to Example~\ref{ex:rankone}. For the path of Example~\ref{ex:path} the maximal cliques are $\{1,2\}$, $\{2,3\}$ and $\{3,4\}$. The two edges of the junction tree give the blocks $\Delta_{12,234}$ and $\Delta_{123,34}$, which both have rank one exactly when
\[
\delta_{13}\,\delta_{22}=\delta_{12}\,\delta_{23},\qquad \delta_{14}\,\delta_{22}=\delta_{12}\,\delta_{24},\qquad \delta_{24}\,\delta_{33}=\delta_{23}\,\delta_{34}.
\]
These are binomial equations, and the same binomials in the entries of $\Sigma$ define the model itself (Section~\ref{sec:discussion}). The entries of $\Delta$ at the vertices and edges can be chosen freely and determine all the others. If $\sigma_{12}$ is multiplied by $\delta$, the model is preserved by multiplying $\sigma_{13}$ and $\sigma_{14}$ by $\delta$ as well. If $\sigma_{22}$ is multiplied by $\delta$, it is preserved by dividing $\sigma_{13}$ and $\sigma_{14}$ by $\delta$.
\end{example}

For a disconnected $G$ the theorem applies to each component, with arbitrary $\delta_{ab}$ between components.

\subsection{Ancestral graphs}\label{sec:ancestral}

We now consider ancestral graphs, which contain undirected and directed acyclic graphs as special cases. A mixed graph has at most one edge between two vertices, which is undirected ($\edge{a}{b}$), directed ($a\to b$) or bidirected ($a\leftrightarrow b$). A vertex $a$ is an ancestor of $b$ if $a=b$ or there is a directed path from $a$ to $b$. The graph is \emph{ancestral} if it has no directed cycle, if $a\leftrightarrow b$ implies that neither vertex is an ancestor of the other, and if no arrowhead points at a vertex of an undirected edge.

An interior vertex of a path is a collider if both adjacent edges of the path have an arrowhead at it. A path is m-connecting given $C$ if none of its non-colliders is in $C$ and each of its colliders is an ancestor of a vertex of $C$. Two vertices are m-separated by $C$ if no path between them is m-connecting given $C$. For an undirected graph this is separation, and for a directed acyclic graph it is d-separation. We write $\cI(G)$ for the set of statements $\st{i}{j}{C}$ such that $C$ m-separates $i$ from $j$. Markov and faithful matrices are then defined as in Section~\ref{sec:graphical}.

For $\tau_k>0$ a change of the variance of $Y_k$ replaces $Y_k$ by a noisy copy. On the graph this adds a child of $k$ and marginalizes over $k$ itself. The marginal of an ancestral graph over a set $V$ of its vertices is an ancestral graph on $V$ with the same m-separation statements among $V$ \citep{richardson2002}. A marginal of a directed acyclic graph need not be one, so ancestral graphs are the natural class here. For $K\subseteq[n]$ let $G^+$ be obtained from $G$ by adding a new vertex $k'$ and the edge $k\to k'$ for each $k\in K$. We write $G^{\langle K\rangle}$ for the marginal of $G^+$ over the vertices that are not in $K$, with each $k'$ relabelled as $k$.

For one covariance the sets $\cI(G)^{kl}$ and $\cI_0$ are defined as in Section~\ref{sec:undirected}, with m-separation in place of separation for screened statements. The graph of the inverse of a faithful matrix is the undirected graph $G^a$ on $[n]$ in which two vertices are adjacent when they are joined in $G$ by a path on which every interior vertex is a collider \citep{richardson2002}. It contains the edges of $G$ and equals $G$ when $G$ is undirected. For a directed acyclic graph it is the moral graph, in which two vertices are adjacent when they are adjacent in $G$ or have a common child.

\begin{theorem}\label{thm:anc}
Let $G$ be an ancestral graph.
\begin{enumerate}[label=\textup{(\alph*)},leftmargin=2.4em]
\item Let $K\subseteq[n]$ and $\Sigma'=\Sigma+\sum_{k\in K}\tau_kE_{kk}\in\PD$. If $\Sigma$ is Markov to $G$ then $\Sigma'$ is Markov to $G^{\langle K\rangle}$, while if $\Sigma$ is faithful to $G$ then $\Sigma'$ is faithful to $G^{\langle K\rangle}$ for generic $\tau$.
\item Let $k\neq l$ and $\Sigma'=\Sigma+\tau(E_{kl}+E_{lk})\in\PD$. The conclusions of Theorem~\ref{thm:ugcov} hold, with $(G^a)^{kl}$ in place of $G^{kl}$ for the graph of the inverse.
\end{enumerate}
\end{theorem}

By part (a) each perturbed variable becomes latent and is observed only through a noisy copy, whether its variance increases or decreases. For an undirected graph $G^{\langle K\rangle}$ has the same separation statements as $G^K$. For a directed acyclic graph it is in general a mixed graph, because the children of a perturbed vertex $k$ and its copy $k'$ share the marginalized parent $k$. Its statements need not be those of any directed acyclic graph. For the graph $1\to4\leftarrow3\to5\leftarrow2$, a change of $\sigma_{33}$ gives the edges $1\to4$, $2\to5$, $3\leftrightarrow4$, $3\leftrightarrow5$ and $4\leftrightarrow5$. A directed acyclic graph with these statements would have the same adjacencies, and $\st{1}{5}{\emptyset}$ and $\st{2}{4}{\emptyset}$ would force its edge between $4$ and $5$ to point both ways.

In part (b), the neighbours of a vertex of a directed acyclic graph in $G^a$ are its parents, its children and the other parents of its children. The graph of the inverse gains every edge among $k$, $l$ and these vertices. As for undirected graphs, the resulting structure need not be that of an ancestral graph.

\begin{figure}[t]
\centering
\begin{tikzpicture}[every node/.style={circle,draw,inner sep=1.5pt,minimum size=15pt,font=\small},scale=0.95,>=stealth]
\begin{scope}
\node (1) at (0,0) {1}; \node (2) at (1.3,0) {2}; \node (3) at (2.6,0) {3}; \node (4) at (3.9,0) {4};
\node (2p) at (1.3,1.05) {$2'$};
\draw[->] (1)--(2); \draw[->] (2)--(3); \draw[->] (3)--(4); \draw[->] (2)--(2p);
\node[draw=none] at (1.95,-0.7) {(a)};
\end{scope}
\begin{scope}[xshift=6cm]
\node (1) at (0,0) {1}; \node (2p) at (1.3,0) {$2'$}; \node (3) at (2.6,0) {3}; \node (4) at (3.9,0) {4};
\draw[->] (1)--(2p); \draw[<->] (2p)--(3); \draw[->] (3)--(4);
\draw[->] (1) to[bend left=45] (3);
\node[draw=none] at (1.95,-0.7) {(b)};
\end{scope}
\end{tikzpicture}
\caption{A change of $\sigma_{22}$ in the directed path of Example~\ref{ex:anc}, with (a) the graph $G^+$ and (b) its marginal over $\{1,2',3,4\}$.}
\label{fig:anc}
\end{figure}
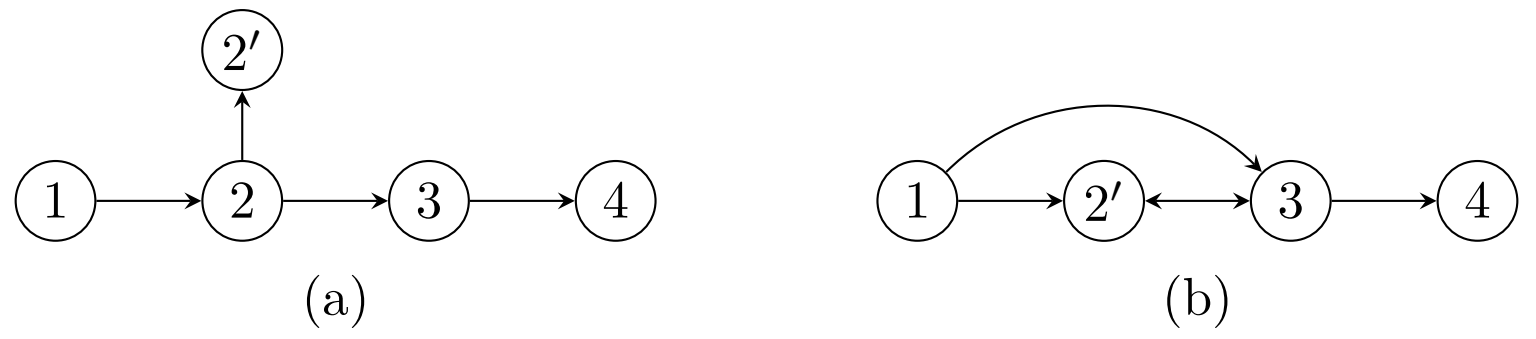

\begin{example}\label{ex:anc}
The matrix of Example~\ref{ex:path} is also faithful to the directed path $1\to2\to3\to4$, which has the same separation statements as the undirected path. For a change of $\sigma_{22}$ the graph $G^+$ has the additional edge $2\to2'$. Its marginal over $\{1,2',3,4\}$ has the edges $1\to2'$, $1\to3$, $2'\leftrightarrow3$ and $3\to4$ (Figure~\ref{fig:anc}). Relabelling $2'$ as $2$ gives $G^{\langle\{2\}\rangle}$. The bidirected edge records that $2'$ and $3$ are both children of the marginalized vertex $2$. The m-separation statements of this graph are the four statements of Example~\ref{ex:pathvar}. Adding the edge $1\to3$ to the path gives a directed acyclic graph with the same statements.

For a change of $\sigma_{23}$ the structure is $\{\st{1}{4}{2},\st{1}{4}{3}\}$ by Example~\ref{ex:pathcov}. No ancestral graph $H$ has exactly these two statements. Indeed, we may assume that $H$ is maximal, so that two vertices are adjacent unless some set m-separates them \citep{richardson2002}. All pairs are then adjacent in $H$ except $\{1,4\}$. If $2$ is a collider on the path through $1$, $2$ and $4$, this path is m-connecting given $\{2\}$. Otherwise it is m-connecting given $\{3\}$. Each case contradicts one of the two statements.
\end{example}

The parameters of a directed acyclic graph that can be changed with no covariation are found as in Corollary~\ref{cor:free}, with moral graphs in the role of the undirected graph. Write $\pa(k)$ for the set of parents of a vertex $k$ and $\chd(k)$ for the set of its children. A set $A$ of vertices is ancestral if it contains the parents of each of its vertices. Its moral graph has vertex set $A$, with two vertices adjacent when they are adjacent in $G$ or have a common child in $A$.

\begin{corollary}\label{cor:dagfree}
Let $\Sigma$ be faithful to a directed acyclic graph $G$ and let $\tau$ be a generic admissible size.
\begin{enumerate}[label=\textup{(\alph*)},leftmargin=2.4em]
\item The matrix $\Sigma+\tau E_{kk}$ is faithful to $G$ if and only if $k$ is simplicial in the moral graph of every ancestral set that contains $k$. This holds exactly when the vertices of $\chd(k)\cup\{k\}$ are pairwise adjacent and $\pa(a)\subseteq\pa(b)$ for every edge $a\to b$ between two of them.
\item The matrix $\Sigma+\tau(E_{kl}+E_{lk})$ is faithful to $G$ if and only if $k$ is adjacent to $l$ and both are simplicial in the moral graph of every ancestral set that contains them.
\end{enumerate}
\end{corollary}

A vertex without children satisfies the condition in (a), since its neighbours in a moral graph are its parents, which the moralization joins. A vertex with a single child $c$ satisfies it exactly when its parents are also parents of $c$. An ancestral set that contains $k$ and $l$ is among those considered in (a) for each of the two vertices. Part (b) therefore holds when two adjacent vertices both satisfy the condition in (a). The converse fails, unlike for undirected graphs. In the graph with the edges $1\to2$, $2\to3$, $2\to4$, $1\to4$ and $3\to4$ the only statement is $\st{1}{3}{2}$, so $\sigma_{24}$ is free although $\sigma_{22}$ is not. The direction of the edges matters, as the next example shows.

\begin{example}\label{ex:collider}
Reverse the last edge of the directed path of Example~\ref{ex:anc}. The graph $1\to2\to3\leftarrow4$ has a collider at $3$, and the matrix
\[
\Sigma=\begin{pmatrix}1&1&1&0\\1&2&2&0\\1&2&4&1\\0&0&1&1\end{pmatrix}
\]
is faithful to it. It is the covariance matrix of $Y_1=\varepsilon_1$, $Y_2=Y_1+\varepsilon_2$, $Y_3=Y_2+Y_4+\varepsilon_3$ and $Y_4=\varepsilon_4$ with independent standard normal errors. Its structure is
\[
\ci{\Sigma}=\{\st{1}{3}{2},\st{1}{3}{24},\st{1}{4}{\emptyset},\st{1}{4}{2},\st{1}{4}{23},\st{2}{4}{\emptyset},\st{2}{4}{1}\}.
\]
The vertices $1$ and $4$ have a single child but no parents, while $3$ has no children. By Corollary~\ref{cor:dagfree} the parameters $\sigma_{11}$, $\sigma_{33}$, $\sigma_{44}$ and $\sigma_{34}$ are free. For the path of Example~\ref{ex:path} only $\sigma_{11}$ and $\sigma_{44}$ are (Example~\ref{ex:free}). The variance $\sigma_{22}$ is not free, because the parent $1$ of vertex $2$ is not a parent of its child $3$. Adding $\tau$ to it gives $\apm{13|2}(\Sigma')=\tau\sigma_{13}=\tau$. The statements $\st{1}{3}{24}$ and $\st{1}{4}{23}$ are lost in the same way. The four statements that remain say that $Y_4$ is independent of $(Y_1,Y_2)$. They are the m-separation statements of the graph $G^{\langle\{2\}\rangle}$ of Theorem~\ref{thm:anc}(a), which has the edges $1\to2$, $1\to3$, $2\leftrightarrow3$ and $4\to3$. A change of $\sigma_{23}$ leaves the same four statements. The moral graph has the edges $\edge{1}{2}$, $\edge{2}{3}$, $\edge{2}{4}$ and $\edge{3}{4}$. Every vertex is thus a neighbour of $2$ or of $3$, so the graph of the inverse becomes complete.
\end{example}

\subsection{Several covariances}\label{sec:several}

Theorems~\ref{thm:ugvar} and~\ref{thm:anc}(a) change several variances at once, but Theorems~\ref{thm:ugcov} and~\ref{thm:anc}(b) change a single covariance. This section treats several covariances changed together. For an ancestral graph the structure stays cancellation-free after each change, so the outcome can again be read from the graph. Let $P$ be a set of pairs of distinct vertices, the positions of the changed covariances, and let $V(P)$ be the set of vertices of these pairs. For a statement $s=\st{i}{j}{C}$ let $P_s$ consist of the pairs of $P$ whose vertices both occur in $s$, either as endpoints or in $C$.

The rule is read in an undirected graph $G_s$ attached to each statement $s$ of an ancestral graph $G$. A vertex is \emph{anterior} to $s$ if it occurs in $s$ or is the first vertex of a path that ends at a vertex of $s$ and whose edges are undirected or directed towards that end. The vertices of $G_s$ are those anterior to $s$. Two of them are adjacent in $G_s$ if they are joined in $G$ by a path through such vertices on which every interior vertex is a collider. In particular, vertices adjacent in $G$ are adjacent in $G_s$. A set $C$ m-separates $i$ from $j$ in $G$ if and only if it separates them in $G_s$ \citep{richardson2002}. If $G$ is undirected, $G_s$ consists of the connected components of $G$ that meet $s$. If $G$ is a directed acyclic graph, $G_s$ is the moral graph of the smallest ancestral set containing $s$ \citep{lauritzen1996}.

Let $\cI(G)^P$ be the set of statements $s=\st{i}{j}{C}$ such that $C\setminus V(P_s)$ separates $i$ from $j$ in $G_s+P_s$, the graph obtained from $G_s$ by adding the pairs of $P_s$ as edges. When all vertices occur in $s$, the graph $G_s$ is the graph $G^a$ of Section~\ref{sec:ancestral}.

\begin{theorem}\label{thm:several}
Let $G$ be an ancestral graph, $P$ a set of pairs and
\[
\Sigma'=\Sigma+\sum_{\{k,l\}\in P}\tau_{kl}(E_{kl}+E_{lk})\in\PD.
\]
If $\Sigma$ is Markov to $G$ then $\cI(G)^P\subseteq\ci{\Sigma'}$. If $\Sigma$ is faithful to $G$, the following hold for generic $\tau$.
\begin{enumerate}[label=\textup{(\alph*)},leftmargin=2.4em]
\item $\ci{\Sigma'}=\cI(G)^P$.
\item The graph of the non-zero off-diagonal entries of $(\Sigma')^{-1}$ is $(G^a+P)^{V(P)}$.
\end{enumerate}
\end{theorem}

A changed covariance that lies inside a statement acts as an additional edge, and its two vertices can no longer be used to separate. A pair of $P$ with a vertex outside the statement plays no role, because the corresponding minor does not contain the changed entry. For a single pair the theorem reduces to the rules of Theorems~\ref{thm:ugcov} and~\ref{thm:anc}(b) for one covariance. The first claim holds for all sizes because a statement of $\cI(G)^P$ follows from the rank of a submatrix of $\Sigma$ without changed entries.

For an undirected graph the rule can be read in $G$ itself. A statement survives when $C\setminus V(P_s)$ separates $i$ from $j$ in $G+P_s$. For a directed acyclic graph the pairs of $P_s$ are added to a moral graph, which depends on the statement through the ancestors of its vertices.

The theorem describes how perturbations compose. The covariances of $P$ can be changed one at a time in any order, with the same final structure. This structure is contained in the intersection of those obtained from each change alone, and it can be strictly smaller. Changes that preserve the structure can be combined, so free parameters can be changed together and the matrix stays faithful to $G$ for generic sizes. Other variances can be included as well, since a change of variances leaves the matrix faithful to an ancestral graph, to which the theorem applies.

\begin{example}\label{ex:several}
In Example~\ref{ex:path} let $P=\{\{1,2\},\{3,4\}\}$. By Theorem~\ref{thm:ugcov} a change of $\sigma_{12}$ alone leaves the statements $\st{1}{4}{3}$, $\st{1}{4}{23}$, $\st{2}{4}{3}$ and $\st{2}{4}{13}$. A change of $\sigma_{34}$ alone leaves $\st{1}{3}{2}$, $\st{1}{3}{24}$, $\st{1}{4}{2}$ and $\st{1}{4}{23}$. The statement $\st{1}{4}{23}$ survives either change but not both. All four vertices occur in it, so $P_s=P$ and the set $C\setminus V(P_s)$ is empty, while $G+P=G$ is connected. Each of the other six statements contains one of the two pairs, and for each of them $C\setminus V(P_s)$ is empty as well. Hence $\ci{\Sigma'}=\emptyset$ for generic sizes, and the graph of the inverse is complete. The directed path of Example~\ref{ex:anc} leads to the same conclusion, because its graphs $G_s$ are paths as well. In Example~\ref{ex:kite} the parameters $\sigma_{11}$, $\sigma_{22}$, $\sigma_{33}$ and $\sigma_{13}$ are free by Example~\ref{ex:free}. Changing all four at once leaves $\ci{\Sigma}$ unchanged.
\end{example}

\section{Sensitivity of the examination marks model}\label{sec:application}

\begin{figure}[t]
\centering
\begin{minipage}[c]{0.4\textwidth}
\centering
\begin{tikzpicture}[every node/.style={circle,draw,inner sep=1.5pt,minimum size=15pt,font=\small},scale=0.95]
\node (1) at (-1.3,0.8) {1}; \node (2) at (-1.3,-0.8) {2}; \node (3) at (0,0) {3};
\node (4) at (1.3,0.8) {4}; \node (5) at (1.3,-0.8) {5};
\draw (1)--(2) (1)--(3) (2)--(3) (3)--(4) (3)--(5) (4)--(5);
\node[draw=none] at (-1.3,1.35) {mechanics}; \node[draw=none] at (-1.3,-1.35) {vectors};
\node[draw=none] at (0,0.62) {algebra};
\node[draw=none] at (1.3,1.35) {analysis}; \node[draw=none] at (1.3,-1.35) {statistics};
\end{tikzpicture}
\end{minipage}%
\begin{minipage}[c]{0.56\textwidth}
\centering
$\hat\Sigma=\begin{pmatrix}
302.3&125.8&100.4&99.7&108.4\\
125.8&170.9&84.2&83.6&90.9\\
100.4&84.2&111.6&110.8&120.5\\
99.7&83.6&110.8&217.9&153.8\\
108.4&90.9&120.5&153.8&294.4
\end{pmatrix}$
\end{minipage}
\caption{The graph selected by \citet{whittaker1990} for the marks of $88$ students in five examinations, and the maximum likelihood estimate $\hat\Sigma$ in its model, rounded to one decimal.}
\label{fig:marks}
\end{figure}

The results are now applied to a covariance matrix estimated from data, whose parameters are known only up to sampling error. Their standard errors give a natural scale for the size of a change. \citet{mardia1979} report the marks of $88$ students in five examinations. These are mechanics, vectors, algebra, analysis and statistics, numbered from $1$ to $5$ in this order. \citet{whittaker1990} selected for these data the graph $G$ of Figure~\ref{fig:marks}, in which algebra separates the first two examinations from the last two. Its model is the set of matrices that satisfy ${\{1,2\}\indep\{4,5\}\,|\,3}$. The maximum likelihood estimate $\hat\Sigma$ in this model \citep{lauritzen1996}, shown in Figure~\ref{fig:marks}, agrees with the sample covariance matrix on the cliques $\{1,2,3\}$ and $\{3,4,5\}$. Its other entries are $\hat\sigma_{ab}=\hat\sigma_{a3}\hat\sigma_{3b}/\hat\sigma_{33}$. All off-diagonal entries of $\hat\Sigma^{-1}$ are non-positive, so the matrix is faithful to $G$ \citep{fallat2017}. Its structure consists of the sixteen statements $\st{a}{b}{C}$ in which $a$ is $1$ or $2$, $b$ is $4$ or $5$, and the conditioning set $C$ contains the vertex $3$.

The model has eleven parameters, the five variances and the six covariances on the edges. Under normality the standard error of an estimated covariance is approximately $((\sigma_{ii}\sigma_{jj}+\sigma_{ij}^2)/n)^{1/2}$ with $n=88$. This is $15$ per cent of the estimate for each variance. For the covariances it ranges from $18$ to $22$ per cent. In relative terms the least precise parameter is the covariance of mechanics and algebra, estimated as $\hat\sigma_{13}=100.4$ with a standard error of $22.3$. By Corollary~\ref{cor:free} six parameters can be varied with no covariation. All vertices but $3$ are simplicial. The adjacent pairs among them are $\{1,2\}$ and $\{4,5\}$. The free parameters are thus $\sigma_{11}$, $\sigma_{22}$, $\sigma_{44}$, $\sigma_{55}$, $\sigma_{12}$ and $\sigma_{45}$. By Section~\ref{sec:several} they can be changed together, and the sixteen statements remain when each of the six is moved by one standard error. The five parameters that involve algebra are not free.

Suppose that $\sigma_{13}$ is increased by one standard error, $\tau=22.3$. By Theorem~\ref{thm:ugcov} the perturbed matrix satisfies only the four statements $\st{2}{4}{3}$, $\st{2}{4}{35}$, $\st{2}{5}{3}$ and $\st{2}{5}{34}$, in which mechanics does not occur. The graph of its inverse is complete. Proposition~\ref{thm:cov}(b) gives $\apm{14|3}(\Sigma')=-\tau\hat\sigma_{34}$ and $\apm{15|3}(\Sigma')=-\tau\hat\sigma_{35}$. Given algebra, the partial correlation of mechanics becomes $-0.165$ with analysis and $-0.145$ with statistics. For comparison, the four sample partial correlations given algebra between the two groups of examinations are at most $0.093$. In this respect the perturbed matrix is further from the model than the sample covariance matrix itself. The covariation of Theorem~\ref{thm:rankone} that keeps the model multiplies $\sigma_{14}$ and $\sigma_{15}$ by the same factor as $\sigma_{13}$, here $1.22$, which fills the first row of the block $\Delta_{123,345}$.

A change of $\sigma_{33}$ affects every statement. The graph $G^{\{3\}}$ is complete, so by Theorem~\ref{thm:ugvar} none of the sixteen statements survives. Identity \eqref{eq:var} gives $\apm{ab|3}(\Sigma')=\tau\hat\sigma_{ab}$ for $a\in\{1,2\}$ and $b\in\{4,5\}$. With an increase of one standard error, $\tau=16.8$, the four partial correlations given algebra lie between $0.07$ and $0.09$. With a decrease of the same size they lie between $-0.16$ and $-0.12$. A noisy algebra mark thus fails to separate the two groups. Theorem~\ref{thm:several} covers joint changes. If the three parameters of mechanics, $\sigma_{11}$, $\sigma_{12}$ and $\sigma_{13}$, are changed together, the structure is again that of ${2\indep\{4,5\}\,|\,3}$, since no changed position lies inside a statement that omits mechanics. A change of $\sigma_{13}$ and $\sigma_{34}$ leaves only $\st{2}{5}{3}$.

\section{Discussion}\label{sec:discussion}

A Schur multiplier preserves the vanishing of an almost-principal minor of a symmetric matrix for structural reasons only through a block of rank one, and the same blocks along a junction tree describe the multipliers that preserve a decomposable graphical model. Beyond these cases the almost-principal minors of a perturbed covariance matrix are explicit polynomials in the size of the perturbation, which determines its conditional independence structure. For graphical models a change of variances is an operation on the graph, whereas a change of covariances yields a structure that is read from the graph by a separation rule but need not be the structure of any graph.

Covariation schemes were introduced to keep a perturbed covariance matrix inside a given model \citep{gorgen2020}. They fill a row, a column or the whole of the relevant block of the multiplier with the same factor. Theorem~\ref{thm:rankone} shows that these are instances of one condition, a block of rank one, and that nothing else works for a single statement. Standard sensitivity analyses compare the original and the perturbed distribution through a divergence, which does not show whether the latter still belongs to the model. In general it does not, and our results give its conditional independence structure. In the examination marks, a change of one standard error in a single covariance induces partial correlations between the two groups that are larger in absolute value than those observed in the sample.

The results also connect with the algebraic description of Gaussian graphical models. By Theorem~\ref{thm:decomposable} the model-preserving multipliers of a decomposable graph are defined by binomial equations, the $2\times2$ minors of the blocks. When every separator is a single vertex, as for trees, the same binomials in the entries of $\Sigma$ define the model, whose vanishing ideal is then toric \citep{misra2021}. For larger separators the model needs minors of higher order, and the relation between these equations and those of the multipliers remains an open question.

From the point of view of matrix theory, the results concern the pattern of vanishing principal and almost-principal minors of a symmetric matrix \citep{brualdi2012,fallat2018} under the simplest perturbations. A change of a diagonal entry acts on the graph of the inverse by joining neighbours, as in one step of an elimination, while keeping the vertex. A change of an off-diagonal entry moves along the positive definite completions of a partial matrix \citep{massa2017}, and the relations that survive are those forced by the two principal submatrices that the completions share. For several such entries, a relation survives only when the rank of an unchanged submatrix forces it.

Three limitations remain. First, after a change of covariances the rules give the whole structure only for faithful matrices. A matrix that is only Markov to the graph keeps the statements given by the rules, but it may keep others too. Without a graph, the rule needs a cancellation-free structure, and otherwise the outcome depends on the partial covariances. Second, a model can impose equations that are not conditional independence statements. In Example~\ref{ex:kite} no statement survives a change of $\sigma_{44}$, yet the tetrad $\sigma_{21}\sigma_{43}-\sigma_{41}\sigma_{23}=0$ holds on the whole model and does not involve $\sigma_{44}$. The perturbed matrices thus form a smaller set than those satisfying the surviving statements, which we have not characterized. Third, model-preserving multipliers are characterized only for decomposable graphs. For other graphs, rank-one blocks for the defining statements are sufficient, but their necessity is open.

\section*{Funding}
This work was funded by grant PID2023-153222OB-I00 of the Agencia Estatal de Investigaci\'on.

\section*{Statement on AI use}
The author used the generative AI tool Claude (Anthropic) in preparing this article. The tool assisted in the formulation of the results, the development of the proofs, their verification and the polishing of their presentation. It also supported writing, exposition and proofreading. The author checked all the proofs and takes full responsibility for the content.

\bibliographystyle{plainnat}
\bibliography{schur_minors_arxiv}

@article{boege2019,
  author = {Boege, T. and D'Al\`i, A. and Kahle, T. and Sturmfels, B.},
  title = {The geometry of gaussoids},
  journal = {Found. Comput. Math.},
  volume = {19},
  year = {2019},
  pages = {775--812}
}

@article{brualdi2012,
  author = {Brualdi, R. A. and Deaett, L. and Olesky, D. D. and van den Driessche, P.},
  title = {The principal rank characteristic sequence of a real symmetric matrix},
  journal = {Linear Algebra Appl.},
  volume = {436},
  year = {2012},
  pages = {2137--2155}
}

@article{castillo2003,
  author = {Castillo, E. and Kj{\ae}rulff, U.},
  title = {Sensitivity analysis in {G}aussian {B}ayesian networks using a symbolic-numerical technique},
  journal = {Reliab. Eng. Syst. Saf.},
  volume = {79},
  year = {2003},
  pages = {139--148}
}

@book{drton2009,
  author = {Drton, M. and Sturmfels, B. and Sullivant, S.},
  title = {Lectures on Algebraic Statistics},
  series = {Oberwolfach Seminars},
  volume = {39},
  publisher = {Birkh\"auser},
  address = {Basel},
  year = {2009}
}

@article{drton2020,
  author = {Drton, M. and Robeva, E. and Weihs, L.},
  title = {Nested covariance determinants and restricted trek separation in {G}aussian graphical models},
  journal = {Bernoulli},
  volume = {26},
  year = {2020},
  pages = {2503--2540}
}

@article{fallat2018,
  author = {Fallat, S. M. and Mart\'inez-Rivera, X.},
  title = {The quasi principal rank characteristic sequence},
  journal = {Linear Algebra Appl.},
  volume = {548},
  year = {2018},
  pages = {42--56}
}

@article{gomez2008,
  author = {G\'omez-Villegas, M. A. and Main, P. and Susi, R.},
  title = {Extreme inaccuracies in {G}aussian {B}ayesian networks},
  journal = {J. Multivariate Anal.},
  volume = {99},
  year = {2008},
  pages = {1929--1940}
}

@article{gorgen2020,
  author = {G\"orgen, C. and Leonelli, M.},
  title = {Model-preserving sensitivity analysis for families of {G}aussian distributions},
  journal = {J. Mach. Learn. Res.},
  volume = {21},
  year = {2020},
  pages = {1--32}
}

@book{horn2013,
  author = {Horn, R. A. and Johnson, C. R.},
  title = {Matrix Analysis},
  edition = {Second},
  publisher = {Cambridge University Press},
  address = {Cambridge},
  year = {2013}
}

@book{lauritzen1996,
  author = {Lauritzen, S. L.},
  title = {Graphical Models},
  publisher = {Oxford University Press},
  address = {Oxford},
  year = {1996}
}

@article{lnenicka2007,
  author = {Ln\v{e}ni\v{c}ka, R. and Mat\'u\v{s}, F.},
  title = {On {G}aussian conditional independence structures},
  journal = {Kybernetika (Prague)},
  volume = {43},
  year = {2007},
  pages = {327--342}
}

@book{mardia1979,
  author = {Mardia, K. V. and Kent, J. T. and Bibby, J. M.},
  title = {Multivariate Analysis},
  publisher = {Academic Press},
  address = {London},
  year = {1979}
}

@article{massa2017,
  author = {Massa, M. S. and Riccomagno, E.},
  title = {Algebraic representations of {G}aussian {M}arkov combinations},
  journal = {Bernoulli},
  volume = {23},
  year = {2017},
  pages = {626--644}
}

@article{richardson2002,
  author = {Richardson, T. and Spirtes, P.},
  title = {Ancestral graph {M}arkov models},
  journal = {Ann. Statist.},
  volume = {30},
  year = {2002},
  pages = {962--1030}
}

@article{sadeghi2014,
  author = {Sadeghi, K. and Lauritzen, S. L.},
  title = {Markov properties for mixed graphs},
  journal = {Bernoulli},
  volume = {20},
  year = {2014},
  pages = {676--696}
}

@article{styan1973,
  author = {Styan, G. P. H.},
  title = {Hadamard products and multivariate statistical analysis},
  journal = {Linear Algebra Appl.},
  volume = {6},
  year = {1973},
  pages = {217--240}
}

@article{sullivant2010,
  author = {Sullivant, S. and Talaska, K. and Draisma, J.},
  title = {Trek separation for {G}aussian graphical models},
  journal = {Ann. Statist.},
  volume = {38},
  year = {2010},
  pages = {1665--1685}
}

@book{whittaker1990,
  author = {Whittaker, J.},
  title = {Graphical Models in Applied Multivariate Statistics},
  publisher = {Wiley},
  address = {Chichester},
  year = {1990}
}

@article{fallat2017,
  author = {Fallat, S. and Lauritzen, S. and Sadeghi, K. and Uhler, C. and Wermuth, N. and Zwiernik, P.},
  title = {Total positivity in {M}arkov structures},
  journal = {Ann. Statist.},
  volume = {45},
  year = {2017},
  pages = {1152--1184}
}

@article{lin2014,
  author = {Lin, S. and Uhler, C. and Sturmfels, B. and B\"uhlmann, P.},
  title = {Hypersurfaces and their singularities in partial correlation testing},
  journal = {Found. Comput. Math.},
  volume = {14},
  year = {2014},
  pages = {1079--1116}
}

@article{saeed2020,
  author = {Saeed, B. and Belyaeva, A. and Wang, Y. and Uhler, C.},
  title = {Anchored causal inference in the presence of measurement error},
  journal = {Proc. Mach. Learn. Res.},
  volume = {124},
  year = {2020},
  pages = {619--628}
}

@article{misra2021,
  author = {Misra, P. and Sullivant, S.},
  title = {Gaussian graphical models with toric vanishing ideals},
  journal = {Ann. Inst. Statist. Math.},
  volume = {73},
  year = {2021},
  pages = {757--785}
}

\appendix

\section{Proofs}\label{app:proofs}

Two properties of Gaussian vectors are used throughout. First, if $A\indep B\,|\,D$ holds for $\Sigma$, then $\st{a}{b}{D\cup A'\cup B'}\in\ci{\Sigma}$ for all $a\in A$, $b\in B$, $A'\subseteq A\setminus\{a\}$ and $B'\subseteq B\setminus\{b\}$ \citep{drton2009}. Second, for $k\notin D\cup\{i,j\}$ the partial covariances given $D$ and $D\cup\{k\}$ satisfy the recursion
\begin{equation}\label{eq:recursion}
\sigma_{ij\cdot Dk}=\sigma_{ij\cdot D}-\frac{\sigma_{ik\cdot D}\,\sigma_{jk\cdot D}}{\sigma_{kk\cdot D}},
\end{equation}
where $\sigma_{kk\cdot D}=\det\Sigma_{Dk}/\det\Sigma_D>0$ is the partial variance of $Y_k$ given $Y_D$.

\subsection{Proof of Theorem~\ref{thm:rankone}}\label{app:rankone}

Throughout, $s=\st{i}{j}{C}$ with $C\neq\emptyset$. Expanding the minor along the row of $i$ and the column of $j$ isolates the entry $\sigma_{ij}$. The other entries of this row and column are $u=(\sigma_{ic})_{c\in C}$ and $v=(\sigma_{cj})_{c\in C}$, so that
\begin{equation}\label{eq:expand}
\apm{s}(\Sigma)=\sigma_{ij}\det\Sigma_C-u^\top\adj(\Sigma_C)\,v,
\end{equation}
where $\adj$ is the adjugate. The multiplier rescales $u$ by $D_u=\diag(\delta_{ic})_{c\in C}$ and $v$ by $D_v=\diag(\delta_{cj})_{c\in C}$.

(a) If $\Delta_{iC,jC}=ab^\top$ then $(\Delta\circ\Sigma)_{iC,jC}=\diag(a)\,\Sigma_{iC,jC}\,\diag(b)$. Hence $\apm{s}(\Delta\circ\Sigma)$ is $\apm{s}(\Sigma)$ times the product of the entries of $a$ and $b$, which is not zero.

(b) Let $\cM_s$ be the set of $\Sigma\in\PD$ with $\apm{s}(\Sigma)=0$. Since $\det\Sigma_C>0$, by \eqref{eq:expand} the entry $\sigma_{ij}$ of a matrix in $\cM_s$ is a rational function of its other entries. A relatively open subset of $\cM_s\cap U_\Delta$ thus corresponds to an open set of values of these entries. The matrices of $\cM_s\cap U_\Delta$ for which $\Delta\circ\Sigma$ satisfies the statement are the zeros of $\apm{s}(\Delta\circ\Sigma)$, which form a relatively closed set. Suppose that this set has an interior point. Expanding $\apm{s}(\Delta\circ\Sigma)$ as in \eqref{eq:expand} and substituting $\sigma_{ij}$ gives
\[
\delta_{ij}\det(\Delta_C\circ\Sigma_C)\;u^\top\adj(\Sigma_C)\,v=\det\Sigma_C\;u^\top D_u\adj(\Delta_C\circ\Sigma_C)\,D_v\,v
\]
on an open set of values of $u$, $v$ and $\Sigma_C$. Both sides are polynomials, so the equality holds everywhere. The matrices of the two bilinear forms in $u$ and $v$ therefore coincide. Put a symmetric matrix $X$ with non-zero entries in place of $\Sigma_C$, such that $X$ and $\Delta_C\circ X$ are invertible. The relation $\adj(M)=\det(M)\,M^{-1}$ then turns the equality of these matrices into $\delta_{ij}X^{-1}=D_u(\Delta_C\circ X)^{-1}D_v$. Inverting gives $\delta_{ij}\,\Delta_C\circ X=D_vXD_u$, whose entry in position $(c,d)$ yields $\delta_{ij}\delta_{cd}=\delta_{cj}\delta_{id}$. Every entry of $\Delta_{iC,jC}$ is thus the product of an entry in the column of $j$ and one in the row of $i$, divided by $\delta_{ij}$. The block has rank one, contrary to the assumption. The set therefore has empty interior, so its complement is a dense open subset of the matrices of $U_\Delta$ that satisfy the statement.

For the last claim, recall that $A\indep B\,|\,C$ holds for $\Sigma$ if and only if $\st{i}{j}{C}$ does for all $i\in A$ and $j\in B$. If the block $\Delta_{A\cup C,B\cup C}$ has rank one, so has each $\Delta_{iC,jC}$, and part (a) applies. Conversely, if every $\Delta_{iC,jC}$ has rank one, each entry of the block equals $\delta_{xc_0}\delta_{c_0y}/\delta_{c_0c_0}$ for a fixed $c_0\in C$, so the block has rank one. A block of larger rank thus contains some $\Delta_{iC,jC}$ of rank larger than one. Consider the matrices $\Sigma$ in which the variables outside $\{i,j\}\cup C$ are independent of each other and of the remaining ones. Such a matrix satisfies $A\indep B\,|\,C$ as soon as it satisfies $\st{i}{j}{C}$. By part (b), applied to the principal submatrix indexed by $\{i,j\}\cup C$, some of these matrices are mapped to a matrix that violates $\st{i}{j}{C}$.\qed

\subsection{Proofs of Propositions~\ref{thm:var}, \ref{thm:cov} and~\ref{thm:rule}}\label{app:entries}

The three propositions share one argument, an expansion of the minor $\apm{s}(\Sigma')$ along the changed entries, so they are proved together. The expansion gives identity \eqref{eq:var} for Proposition~\ref{thm:var} and identity \eqref{eq:cov} for Proposition~\ref{thm:cov}. Proposition~\ref{thm:rule} then combines \eqref{eq:cov} with the recursion \eqref{eq:recursion}.

\begin{proof}[Proof of Proposition~\ref{thm:var}]
The perturbation adds $\tau_k$ to the entry $(k,k)$ of $\Sigma_{iC,jC}$ for each $k\in K\cap C$. Expanding the determinant by multilinearity in these rows gives one term for each $T\subseteq K\cap C$, in which the rows in $T$ are replaced by $\tau_k$ times a unit vector. The Laplace expansion along them deletes the rows and columns in $T$. No sign appears, because the same indices are deleted on both sides, which gives \eqref{eq:var}. If $\st{i}{j}{C}$ belongs to the set \eqref{eq:varstruct}, all the coefficients vanish and the statement holds for $\Sigma'$. Otherwise \eqref{eq:var} is a non-zero polynomial, because its monomials in $\tau$ are distinct, so the statement fails for generic $\tau$. As there are finitely many statements, the last claim follows.
\end{proof}

\begin{proof}[Proof of Proposition~\ref{thm:cov}]
If $k$, $l$ do not both occur in $s$, the block $\Sigma_{iC,jC}$ contains neither position $(k,l)$ nor $(l,k)$. In cases (a) and (b) it contains exactly one of them. The identity is then the expansion along that entry. In case (c) it contains both. The coefficient of $\tau$ is the sum of the two cofactors,
\[
L_s=\det\Sigma_{ikD,\,ljD}+\det\Sigma_{ilD,\,kjD},
\]
with rows and columns taken in the order written. The coefficient of $\tau^2$ is minus the minor obtained by deleting the rows and columns $k$, $l$. Taking Schur complements with respect to $\Sigma_D$ turns each cofactor into $\det\Sigma_D$ times a $2\times2$ determinant of partial covariances, and the sum is \eqref{eq:L}. Identity \eqref{eq:covstruct} follows, because the polynomial \eqref{eq:cov} is identically zero exactly when $s\in\ci{\Sigma}$ and $L_s(\Sigma)=Q_s(\Sigma)=0$, while otherwise $s$ fails for generic $\tau$.
\end{proof}

\begin{proof}[Proof of Proposition~\ref{thm:rule}]
A statement of $\ci{\Sigma}$ that does not contain both $k$ and $l$ is untouched. Its minor is the same for $\Sigma$ as for $\Sigma'$. Let $s=\st{i}{j}{C}\in\ci{\Sigma}$ be a statement in which both occur. By \eqref{eq:cov} it holds for $\Sigma'$ at every size if $L_s(\Sigma)=Q_s(\Sigma)=0$, and fails for generic $\tau$ otherwise. We compare this condition with being screened for the three types of Proposition~\ref{thm:cov}.

\emph{Type (a).} The statement has no endpoint outside $\{k,l\}$, so it is not screened, and $L_s=\det\Sigma_C\neq0$.

\emph{Type (b).} Let $s=\st{k}{j}{C}$ with $l\in C$, so that $D=C\setminus\{l\}$. Here $Q_s=0$, and $L_s=0$ means that $\st{l}{j}{D}$ holds. By the recursion \eqref{eq:recursion} the statements $\st{k}{j}{C}$ and $\st{l}{j}{D}$ both hold if and only if $\st{j}{k}{D}$ and $\st{j}{l}{D}$ both hold, that is, when the statement is screened by $j$.

\emph{Type (c).} Write $x=\sigma_{ij\cdot D}$, $a=\sigma_{ik\cdot D}$, $b=\sigma_{il\cdot D}$, $c=\sigma_{jk\cdot D}$, $d=\sigma_{jl\cdot D}$, $p=\sigma_{kk\cdot D}$, $q=\sigma_{ll\cdot D}$ and $r=\sigma_{kl\cdot D}$. By Proposition~\ref{thm:cov} the conditions $L_s=Q_s=0$ amount to $x=0$ and $ad+bc=0$. If it is screened by $i$ then $x=a=b=0$, and if by $j$ then $x=c=d=0$. Both conditions then hold.

This proves the first claim. It also shows that, for generic $\tau$, a statement of type (a) or (b) holds for $\Sigma'$ only if it is screened. The statements that fail for $\Sigma$ fail for $\Sigma'$ for generic $\tau$ by Proposition~\ref{thm:cov}. It remains to show that, if $\ci{\Sigma}$ is cancellation-free, a statement of type (c) with $x=0$ and $ad+bc=0$ is screened. Since $s$ holds for $\Sigma$ and $x=0$, two applications of the recursion \eqref{eq:recursion} give
\begin{equation}\label{eq:constraint}
acq-(ad+bc)r+bdp=0 .
\end{equation}
The statements $\st{i}{j}{D}$ and $\st{i}{j}{D\cup\{k,l\}}$ both hold for $\Sigma$, so one of $a,b,c,d$ vanishes because $\ci{\Sigma}$ is cancellation-free. By symmetry in $i,j$ and in $k,l$ we may assume $a=0$. Then $bc=0$ and \eqref{eq:constraint} reduces to $bdp=0$. If $b=0$ it is screened by $i$. Otherwise $c=d=0$ and it is screened by $j$.
\end{proof}

\subsection{Proofs for undirected graphs}\label{app:undirected}

\begin{proof}[Proof of Theorem~\ref{thm:ugvar}]
Let $i\neq j$ and $C\subseteq[n]\setminus\{i,j\}$. We first show that $C$ separates $i$ from $j$ in $G^K$ if and only if $C\setminus K$ separates them in $G$. Take a path from $i$ to $j$ in $G$ that avoids $C\setminus K$ and delete its vertices in $C\cap K$. Consecutive remaining vertices are joined in $G$ by a path with interior in $K$, so they are adjacent in $G^K$. This gives a path in $G^K$ that avoids $C$. Conversely, replace each edge of a path in $G^K$ that avoids $C$ by a path in $G$ with interior in $K$. The resulting walk avoids $C\setminus K$, since the inserted paths have interior in $K$. Separation by a set implies separation by every larger set. Hence $C\setminus K$ separates $i$ from $j$ in $G$ if and only if $\st{i}{j}{C\setminus T}\in\cI(G)$ for all $T\subseteq K\cap C$. By the first step this is the case exactly when $\st{i}{j}{C}\in\cI(G^K)$. If $\Sigma$ is Markov to $G$, the coefficients in \eqref{eq:var} vanish for the statements of $\cI(G^K)$, which thus hold for $\Sigma'$. If $\Sigma$ is faithful to $G$, the set \eqref{eq:varstruct} is $\cI(G^K)$ and Proposition~\ref{thm:var} applies.
\end{proof}

\begin{proof}[Proof of Theorem~\ref{thm:ugcov}]
If $\Sigma$ is Markov to $G$ then $\cI(G)\subseteq\ci{\Sigma}$, hence $\cI(G)^{kl}\subseteq\ci{\Sigma}^{kl}$, because both properties only ask that certain statements lie in the set. Proposition~\ref{thm:rule} then gives the first claim.

(a) The set $\cI(G)$ is cancellation-free. Indeed, if $D$ separates $i$ from $j$ then $k$ lies in the connected component of at most one of them after deleting $D$, so $D$ separates $k$ from $i$ or from $j$. Proposition~\ref{thm:rule} therefore applies to a faithful matrix, for which $\ci{\Sigma}=\cI(G)$.

(b) The vertices $i$ and $j$ are non-adjacent in the graph of $(\Sigma')^{-1}$ if and only if $\st{i}{j}{C}\in\cI(G)^{kl}$ for $C=[n]\setminus\{i,j\}$. Here $D=[n]\setminus\{i,j,k,l\}$, and $D$ separates a vertex $e$ from the rest of $\{i,j,k,l\}$ if and only if $e$ is adjacent to none of them in $G$. The pair $\{k,l\}$ is of type (a) and becomes adjacent. A pair $\{k,j\}$ with $j\neq l$ stays non-adjacent if and only if $j$ is adjacent to neither $k$ nor $l$ in $G$. A pair $\{i,j\}$ disjoint from $\{k,l\}$ stays non-adjacent if and only if it is non-adjacent in $G$ and one of $i,j$ has no neighbour in $\{k,l\}$. These are the non-adjacencies of $G^{kl}$.

For the last claim, $\cI(G)^{kl}$ contains $\cI_0$ and is the structure of the positive definite matrix $\Sigma'$ by part (a). Let $\Sigma''\in\PD$ with $\cI_0\subseteq\ci{\Sigma''}$ and let $\st{i}{j}{C}\in\cI(G)^{kl}$ be screened by $e$. The statements $\st{e}{a}{D}$ with $a\in\{i,j,k,l\}\setminus\{e\}$ belong to $\cI(G)$. Since $k$ and $l$ do not both occur in any of them, they hold for $\Sigma''$. Hence $Y_e$ is independent of the other variables indexed by $\{i,j,k,l\}$ given $Y_D$, and the first property recalled at the beginning of this appendix gives $\st{i}{j}{C}\in\ci{\Sigma''}$.
\end{proof}

\begin{proof}[Proof of Corollary~\ref{cor:free}]
An undirected graph is determined by its separation statements. By Theorem~\ref{thm:ugvar} the first claim amounts to $G^{\{k\}}=G$, that is, to $k$ being simplicial. By Theorem~\ref{thm:ugcov}(b) the second claim requires $G^{kl}=G$. Equivalently $N[k]\cup N[l]$ is a clique, that is, $k$ and $l$ are adjacent simplicial vertices.

For sufficiency let $\st{i}{j}{C}\in\cI(G)$ involve both $k$ and $l$. Put $D=C\setminus\{k,l\}$. If a set $C'$ separates $i$ from $j$ and $m\in C'$ is simplicial, then $C'\setminus\{m\}$ separates them as well. Indeed, a path from $i$ to $j$ that avoids $C'\setminus\{m\}$ passes through $m$, whose two neighbours on the path are adjacent, so leaving $m$ out gives a path that avoids $C'$. Hence $D$ separates $i$ from $j$. The adjacent vertices $k$ and $l$ lie outside $D$, hence in one connected component of $G$ after deleting $D$. This component does not contain both $i$ and $j$. An endpoint outside $\{k,l\}$ is therefore separated by $D$ from the other vertices of $\{i,j,k,l\}$, so the statement is screened and $\cI(G)^{kl}=\cI(G)$.
\end{proof}

\subsection{Proof of Theorem~\ref{thm:decomposable}}\label{app:decomposable}

Throughout, $G$ is a connected decomposable graph with junction tree $T$, and $S_e$, $A_e$, $B_e$ are as in Section~\ref{sec:undirected}. We call $\Delta_{A_e\cup S_e,B_e\cup S_e}$ the block of the edge $e$. Three facts about junction trees are used \citep{lauritzen1996}. The maximal cliques that contain a given vertex form a subtree of $T$. Every set $S_e$ is non-empty and separates $A_e$ from $B_e$, while $A_e\cup B_e\cup S_e=[n]$. If $C$ is a leaf of $T$ with neighbour $C'$, the vertices of $C\setminus C'$ lie in no other maximal clique, and deleting them from $G$ gives a connected decomposable graph with junction tree $T-C$.

The proof has five steps. Step 1 shows that $\cM(G)$ is defined by the statements $A_e\indep B_e\,|\,S_e$, which gives sufficiency by Theorem~\ref{thm:rankone}. Necessity is proved by induction on the number of maximal cliques, removing a leaf $C$ of $T$. Step 2 shows that the multiplier restricted to the remaining vertices is model-preserving for the smaller graph. Steps 3 and 4 show that the block of the edge that joins $C$ to $T-C$ has rank one. In particular the rows of the removed vertices are proportional to a row that remains, which is used in Step 5 for the blocks of the other edges.

\emph{Step 1. A matrix $\Sigma\in\PD$ lies in $\cM(G)$ if and only if it satisfies $A_e\indep B_e\,|\,S_e$ for all edges $e$ of $T$.} Necessity is the global Markov property. For sufficiency let $i$ and $j$ be non-adjacent. No maximal clique contains both, so the subtree of the maximal cliques that contain $i$ is disjoint from that of the maximal cliques that contain $j$. Let $e$ be the first edge of the path of $T$ that leads from the first subtree to the second. The two subtrees lie on different sides of $e$, and neither $i$ nor $j$ belongs to $S_e$. Up to exchanging the two sides, $i\in A_e$ and $j\in B_e$. The first property recalled at the beginning of this appendix then gives $\st{i}{j}{[n]\setminus\{i,j\}}\in\ci{\Sigma}$, that is, $(\Sigma^{-1})_{ij}=0$. If every block has rank one, Theorem~\ref{thm:rankone} shows that $\Delta$ preserves each statement $A_e\indep B_e\,|\,S_e$, so it is model-preserving for $\cM(G)$.

Conversely, let $\Delta$ be model-preserving for $\cM(G)$. We show by induction on the number of maximal cliques that every block has rank one. With one maximal clique there is nothing to prove. Otherwise let $C$ be a leaf of $T$ with neighbour $C'$. Put $S=C\cap C'$, $R=C\setminus S$ and $V'=[n]\setminus R$. Let $G'$ be the graph obtained from $G$ by deleting $R$, with junction tree $T'=T-C$. Recall that a block with non-zero entries has rank one exactly when all its $2\times2$ minors vanish.

\emph{Step 2. The multiplier $\Delta_{V'}$ is model-preserving for $\cM(G')$.} Let $\Sigma_1\in\cM(G')$ be such that $\Delta_{V'}\circ\Sigma_1$ is positive definite. Let $\Sigma$ be the matrix in which the variables of $V'$ have covariance $\Sigma_1$, while those of $R$ have unit variance and are independent of all the others. Then $\Sigma\in\cM(G)\cap U_\Delta$, so $\Delta\circ\Sigma\in\cM(G)$. This matrix is block diagonal, hence the inverse of its block $\Delta_{V'}\circ\Sigma_1$ vanishes at the non-edges of $G'$. By the induction hypothesis the blocks of $\Delta_{V'}$ for the edges of $T'$ have rank one.

\emph{Step 3. The block $\Delta_{C,C'}$ has rank one.} Let $r\in R$ and $j\in C'\setminus S$. Consider the matrices that satisfy $\st{r}{j}{S}$ in which the variables outside $\{r,j\}\cup S$ are mutually independent and independent of the remaining ones. They belong to $\cM(G)$, because the vertices of $\{r\}\cup S$ are pairwise adjacent and so are those of $\{j\}\cup S$. Every matrix of $\cM(G)$ satisfies $\st{r}{j}{S}$, since $S$ separates $r$ from $j$. Hence $\Delta$ maps none of these matrices to one that violates this statement. By Theorem~\ref{thm:rankone}(b), applied to the principal submatrix indexed by $\{r,j\}\cup S$, the block $\Delta_{rS,jS}$ has rank one. As in the last part of the proof of Theorem~\ref{thm:rankone}, the block with rows $R\cup S=C$ and columns $(C'\setminus S)\cup S=C'$ then has rank one.

\emph{Step 4. The block of the edge $\{C,C'\}$ has rank one.} This block is $\Delta_{C,V'}$. Fix $s_0\in S$, which indexes both a row and a column of it. The block has rank one as soon as
\begin{equation}\label{eq:leaf}
\delta_{cy}\,\delta_{s_0s_0}=\delta_{cs_0}\,\delta_{s_0y}\qquad\text{for all}\ c\in C\ \text{and}\ y\in V'.
\end{equation}
For $y\in C'$ this holds by Step 3. Let $y\notin C'$, and let $e'$ be the first edge of the path of $T'$ from $C'$ to the nearest maximal clique that contains $y$. The maximal cliques that contain $y$ form a subtree that avoids $C'$, so they all lie beyond $e'$, and $y\notin S_{e'}$ because $S_{e'}\subseteq C'$. Up to transposition, the block of $e'$ in $\Delta_{V'}$ therefore has the vertices of $C'$ among its rows and $y$ among its columns.

Let first $c\in S$ and take $t\in S_{e'}$. The rows $c$, $s_0$ and the columns $y$, $t$ belong to the block of $e'$, which has rank one by Step 2, so $\delta_{cy}\delta_{s_0t}=\delta_{ct}\delta_{s_0y}$. Since $t\in C'$, Step 3 gives $\delta_{ct}\delta_{s_0s_0}=\delta_{cs_0}\delta_{s_0t}$. Multiplying the two equalities and cancelling $\delta_{ct}\delta_{s_0t}$ yields \eqref{eq:leaf}.

Let now $c=r\in R$. We use a matrix of the model in which $Y_r$ depends on the other variables only through $Y_{s_0}$. Let $G''$ be obtained from $G$ by deleting the edges between $r$ and the vertices other than $s_0$. It is connected, because the neighbours of $r$ in $G$ are pairwise adjacent, and $\cM(G'')\subseteq\cM(G)$. Let $N$ be the adjacency matrix of $G''$. For small $\varepsilon>0$ the matrix $\Sigma=(I-\varepsilon N)^{-1}=\sum_{m\geq0}\varepsilon^mN^m$ belongs to $\cM(G'')$ and has positive entries. It also lies in $U_\Delta$, as it tends to the identity matrix when $\varepsilon\to0$. In $G''$ the vertex $s_0$ separates $r$ from all other vertices, so $\sigma_{rx}=\rho\,\sigma_{s_0x}$ for every $x\neq r$, with $\rho=\sigma_{rs_0}/\sigma_{s_0s_0}>0$.

The matrix $\Delta\circ\Sigma$ belongs to $\cM(G)$, in which $S$ separates $r$ from $y$, hence $\apm{ry|S}(\Delta\circ\Sigma)=0$. In the submatrix $(\Delta\circ\Sigma)_{rS,yS}$ the row of $r$ has the entries $\rho\,\delta_{rx}\sigma_{s_0x}$ and the row of $s_0$ has the entries $\delta_{s_0x}\sigma_{s_0x}$, for $x\in\{y\}\cup S$. By Step 3, $\delta_{rx}=\lambda\delta_{s_0x}$ for $x\in S$, where $\lambda=\delta_{rs_0}/\delta_{s_0s_0}$. After subtracting $\lambda\rho$ times the row of $s_0$ from the row of $r$, the only entry of the row of $r$ that can be non-zero is the one in the column of $y$. Expanding the determinant along this row gives
\[
0=\apm{ry|S}(\Delta\circ\Sigma)=\rho\,(\delta_{ry}-\lambda\,\delta_{s_0y})\,\sigma_{s_0y}\,\det(\Delta\circ\Sigma)_S .
\]
Since $\rho$, $\sigma_{s_0y}$ and $\det(\Delta\circ\Sigma)_S$ are positive, $\delta_{ry}=\lambda\,\delta_{s_0y}$, which is \eqref{eq:leaf} for $c=r$.

\emph{Step 5. The block of every other edge has rank one.} Let $e'$ be an edge of $T'$. Up to transposition, its block in $\Delta$ is its block in $\Delta_{V'}$ with the rows of $R$ added, and the latter has rank one by Step 2. By \eqref{eq:leaf} the row of $r\in R$ is, on the columns in $V'$, a multiple of the row of $s_0$. This row already belongs to the block because $s_0\in C'$, so the rank remains equal to one.\qed

\subsection{Proof of Theorem~\ref{thm:anc}}\label{app:ancestral}

Part (a) compares the set \eqref{eq:varstruct} with m-separation in $G^+$. Part (b) follows the proof of Theorem~\ref{thm:ugcov} once $\cI(G)$ is known to be cancellation-free, which is Lemma~\ref{lem:L}. Two facts are used \citep{richardson2002}. For disjoint $D,M\subseteq[n]$ there is an ancestral graph on $V=[n]\setminus(D\cup M)$ whose m-separation statements are the $\st{a}{b}{B}$ with $a,b\in V$, $B\subseteq V$ and $\st{a}{b}{B\cup D}\in\cI(G)$. Moreover, every ancestral graph has a faithful Gaussian distribution.

\begin{lemma}\label{lem:L}
Let $G$ be an ancestral graph. Then the set $\cI(G)$ of its m-separation statements is cancellation-free for every pair $\{k,l\}$ of distinct vertices.
\end{lemma}
\begin{proof}
Let $i,j,k,l$ be distinct. Let $D\subseteq[n]\setminus\{i,j,k,l\}$ be such that $\st{i}{j}{D}$ and $\st{i}{j}{D\cup\{k,l\}}$ belong to $\cI(G)$. Marginalizing and conditioning gives an ancestral graph $H$ on $\{i,j,k,l\}$ whose m-separation statements are the $\st{a}{b}{B}$ with $\st{a}{b}{B\cup D}\in\cI(G)$. In $H$ the set $\emptyset$ m-separates $i$ from $j$, and so does $\{k,l\}$. Suppose that no pair $\{a,b\}$ with $a\in\{i,j\}$ and $b\in\{k,l\}$ is m-separated by $\emptyset$, so that each of these pairs is joined by a path without colliders. The vertices $i$, $j$ are not adjacent, and they are not both adjacent to $k$. Otherwise the path $i$, $k$, $j$ would be m-connecting given $\emptyset$ if $k$ is a non-collider on it, and given $\{k,l\}$ if $k$ is a collider. The same holds for $l$, so no path joining these pairs contains both $i$ and $j$.

\emph{Case 1: $i$ and $k$ are adjacent.} Then $j$ is not adjacent to $k$, and their only possible collider-free path is $j$, $l$, $k$. So $l$ is adjacent to $j$ and to $k$, as a non-collider between them. Hence $i$ is not adjacent to $l$. Their only collider-free path is $i$, $k$, $l$, on which $k$ is a non-collider. The path $i$, $k$, $l$, $j$ then has no collider, which contradicts the m-separation of $i$ and $j$ by $\emptyset$.

\emph{Case 2: $i$ and $k$ are not adjacent.} Their only possible collider-free path is $i$, $l$, $k$, so $l$ is adjacent to both, as a non-collider. Then $j$ is not adjacent to $l$, and likewise $k$ is a non-collider on the path $j$, $k$, $l$. The path $i$, $l$, $k$, $j$ has no collider, again a contradiction.
\end{proof}

\begin{proof}[Proof of Theorem~\ref{thm:anc}]
(a) For a statement $s$ on $[n]$ let $s'$ be obtained by replacing each $k\in K$ that occurs in $s$ by $k'$, and let $\mathcal{J}_K$ be the set \eqref{eq:varstruct} with $\cI(G)$ in place of $\ci{\Sigma}$. By the definition of $G^{\langle K\rangle}$, a statement $s$ belongs to $\cI(G^{\langle K\rangle})$ exactly when $s'\in\cI(G^+)$. We show that this is the case if and only if $s\in\mathcal{J}_K$. The two claims then follow as in the proof of Theorem~\ref{thm:ugvar}. If $\Sigma$ is Markov to $G$, the coefficients in \eqref{eq:var} vanish for every $s\in\mathcal{J}_K$. If $\Sigma$ is faithful, the set \eqref{eq:varstruct} is $\mathcal{J}_K$ and Proposition~\ref{thm:var} applies.

\emph{If $s\notin\mathcal{J}_K$ then $s'\notin\cI(G^+)$.} Let $s=\st{i}{j}{C}$. For some $T\subseteq K\cap C$ a path in $G$ between $i$ and $j$ is m-connecting given $C\setminus T$. Extend it by the edge $i\to i'$ if $i\in K$ and likewise for $j$. Its non-colliders are outside $C\setminus T$, hence outside the conditioning set of $s'$, which contains no vertex of $K$. Each of its colliders is an ancestor of some $c\in C\setminus T$, and $c$ or $c'$ is in the conditioning set of $s'$. The extended path is therefore m-connecting in $G^+$.

\emph{If $s\in\mathcal{J}_K$ then $s'\in\cI(G^+)$.} Take a Gaussian vector faithful to $G^+$. The m-separation statements among the vertices of $[n]$ are the same in $G^+$ and in $G$, so the covariance matrix $\Sigma_0$ of the components indexed by $[n]$ is faithful to $G$. Each $k'$ is m-separated from all other vertices by $k$. Hence $Y_{k'}=\beta_kY_k+\varepsilon_k$ with $\beta_k\neq0$, where the $\varepsilon_k$ are independent of each other and of the components indexed by $[n]$. Up to a rescaling of the coordinates in $K$, which does not affect the structure, the covariance matrix of the components indexed by $([n]\setminus K)\cup\{k':k\in K\}$ is $\Sigma_0+\sum_{k\in K}\tau_kE_{kk}$ with $\tau_k>0$. All coefficients in \eqref{eq:var} vanish for $s$, so $s'$ holds for this vector and $s'\in\cI(G^+)$.

(b) The proof of Theorem~\ref{thm:ugcov} uses the graph in two places only. Its part (a) needs $\cI(G)$ to be cancellation-free for $\{k,l\}$, which holds by Lemma~\ref{lem:L}. Its part (b) uses that, for $D=[n]\setminus\{i,j,k,l\}$, the statements $\st{e}{a}{D}$ with $a\in\{i,j,k,l\}\setminus\{e\}$ all belong to $\cI(G)$ exactly when $e$ is adjacent to none of these vertices in $G$. For an ancestral graph this holds with $G^a$ in place of $G$. Indeed, for a matrix faithful to $G$ both conditions say that the row of $e$ in its inverse vanishes at these positions, and $G^a$ is the graph of this inverse. The rest of that proof, including the minimality of $\cI(G)^{kl}$, applies unchanged with $(G^a)^{kl}$ in place of $G^{kl}$.
\end{proof}

\subsection{Proof of Corollary~\ref{cor:dagfree}}\label{app:dagfree}

Here $G$ is a directed acyclic graph, $M_A$ is the moral graph of an ancestral set $A$ and $A_s$ is the smallest ancestral set that contains the vertices of a statement $s$. Thus $G_s=M_{A_s}$, and $s=\st{i}{j}{C}$ belongs to $\cI(G)$ exactly when $C$ separates $i$ from $j$ in $M_{A_s}$ (Section~\ref{sec:several}). If $A'\subseteq A$ are ancestral sets, then $M_{A'}$ is a subgraph of $M_A$. The property of simplicial vertices shown in the proof of Corollary~\ref{cor:free} is also used.

(a) Let $k$ be simplicial in $M_A$ for every ancestral set $A$ that contains it, and let $s=\st{i}{j}{C}\in\cI(G)$ with $k\in C$. By the property of simplicial vertices $C\setminus\{k\}$ separates $i$ from $j$ in $M_{A_s}$, hence in the subgraph $M_{A_t}$ for $t=\st{i}{j}{C\setminus\{k\}}$, so that $t\in\cI(G)$. Hence the set \eqref{eq:varstruct} is $\cI(G)$.

Conversely, let $u$, $v$ be neighbours of $k$ in some $M_A$ that are not adjacent there. Then $k$ has a child in $A$, since a vertex without children in $A$ has only its parents as neighbours in $M_A$, which are pairwise adjacent. Let $s=\st{u}{v}{C}$ with $C=A\setminus\{u,v\}$ and $t=\st{u}{v}{C\setminus\{k\}}$. Then $A_s=A_t=A$, as $k$ has a child in $A$. Hence $s\in\cI(G)$, while $t\notin\cI(G)$ since $u$, $k$, $v$ is a path in $M_A$, and $s$ is lost by Proposition~\ref{thm:var}.

For the second description, suppose first that the vertices of $\chd(k)\cup\{k\}$ are pairwise adjacent with $\pa(a)\subseteq\pa(b)$ for every edge $a\to b$ between two of them. Let $A$ be an ancestral set that contains $k$ and let $W$ consist of $k$ together with its children in $A$. As $G$ is acyclic, some $b\in W$ is a child of all other vertices of $W$. A neighbour $x\neq b$ of $k$ in $M_A$ lies in $W$ or is a parent of some $a\in W$, and in both cases $x\in\pa(b)$ by the inclusions between the sets of parents. The neighbours of $k$ other than $b$ are thus parents of $b$, hence pairwise adjacent in $M_A$ and adjacent to $b$.

Conversely, let $k$ be simplicial in every such moral graph. Two children $a$, $b$ of $k$ have no common child in the smallest ancestral set $A$ that contains them, since it would be an ancestor of one of them. Being neighbours of $k$ in $M_A$, they are adjacent in $G$. Let now $a\to b$ be an edge between two vertices of $\chd(k)\cup\{k\}$ and $A$ the smallest ancestral set containing $b$. Then $b$ is a child of $k$ without children in $A$. A parent $y\neq k$ of $a$ is a neighbour of $k$ in $M_A$, as a parent of $k$ or of its child $a$. Hence $y$ and $b$ are adjacent in $M_A$, thus in $G$ for lack of a common child in $A$, and $y\in\pa(b)$ as $y$ is an ancestor of $b$.

(b) Let $P=\{\{k,l\}\}$. Suppose that $k$ is adjacent to $l$ and that both are simplicial in $M_A$ for every ancestral set $A$ containing them. Let $s=\st{i}{j}{C}\in\cI(G)$ be a statement in which both occur. Then $G_s+P_s=M_{A_s}$, in which $C$ separates $i$ from $j$. So does $C\setminus\{k,l\}$, by the property of simplicial vertices. The other statements belong to $\cI(G)^P$ if and only if they belong to $\cI(G)$. Hence $\cI(G)^P=\cI(G)$, and $\Sigma'$ is faithful to $G$ by Theorem~\ref{thm:several}(a).

Conversely, let $\Sigma'$ be faithful to $G$. If $k$ were not adjacent to $l$, the parents of one of them would d-separate them \citep{lauritzen1996}, and this statement of type (a) fails for $\Sigma'$ by Proposition~\ref{thm:cov}. Suppose that $k$, say, has neighbours $u$, $v$ in some $M_A$ with $k,l\in A$ that are not adjacent there. Let $C=A\setminus\{u,v\}$ and $s=\st{u}{v}{C}$. Then $s\in\cI(G)$ is a statement in which $k$ and $l$ both occur, with $G_s+P_s=M_A$. The path $u$, $k$, $v$ avoids $C\setminus\{k,l\}$, so $s\notin\cI(G)^P$, which contradicts Theorem~\ref{thm:several}(a).\qed

\subsection{Proof of Theorem~\ref{thm:several}}\label{app:several}

The claim for a Markov matrix follows directly from the rank of an unchanged submatrix. For a faithful matrix the proof changes one covariance at a time and applies Proposition~\ref{thm:rule} at each step. For the step from $P$ to $P'=P\cup\{\{k,l\}\}$, Lemma~\ref{lem:several}(a) shows that $\cI(G)^P$ is cancellation-free, and Lemma~\ref{lem:several}(b) that the statements kept by Proposition~\ref{thm:rule} are those of $\cI(G)^{P'}$. Both lemmas rest on Lemma~\ref{lem:GU}, which collects properties of the graphs $G_s$.

Throughout, $G$ is an ancestral graph, and the definitions of Section~\ref{sec:several} are used with a set $U$ of vertices in place of the indices of a statement. Thus $G_U$ has the vertices anterior to $U$, two of them being adjacent if they are joined in $G$ by a path through such vertices on which every interior vertex is a collider. A vertex anterior to a vertex that is anterior to $U$ is itself anterior to $U$. Recall that $G$ has no directed cycle and that a vertex with an arrowhead has no undirected edge \citep{richardson2002}. Consequently, if a vertex $w$ without undirected edges is anterior to a vertex $b\neq w$, a directed path leads from $w$ to $b$. Indeed, the path leaves $w$ by a directed edge, so the next vertex has an arrowhead, hence no undirected edge, and the same applies to each vertex in turn.

\begin{lemma}\label{lem:GU}
Let $U\subseteq U'$ and $U''$ be sets of vertices.
\begin{enumerate}[label=\textup{(\alph*)},leftmargin=2.4em]
\item If all vertices of $U''$ are anterior to $U'$, then $G_{U''}$ is a subgraph of $G_{U'}$.
\item Let $u$ be a vertex of $G_U$ and $uv$ an edge of $G_{U'}$ that is not an edge of $G_U$. Then some vertex $b\in U'$ is not anterior to $U$ and is joined to $u$ by a path in $G_{U\cup\{b\}}$, none of whose vertices other than $u$ is anterior to $U$.
\item Let $l$ be a vertex with an undirected edge. Then every edge of $G_{U\cup\{l\}}$ that meets a vertex anterior to $U$ has both of its endpoints anterior to $U$.
\item Let $l$ be a vertex without undirected edges that is not anterior to $U'$, and let $U_1,U_2\subseteq U'$. If $x_m$ is adjacent to $l$ in $G_{U_m\cup\{l\}}$ for $m=1,2$, then $x_1=x_2$ or $x_1$ is adjacent to $x_2$ in $G_{U'\cup\{l\}}$.
\end{enumerate}
\end{lemma}
\begin{proof}
(a) The vertices anterior to $U''$ are anterior to $U'$, and a path through the former is a path through the latter, so every edge of $G_{U''}$ is an edge of $G_{U'}$.

(b) Let $\rho$ be a path from $u$ to $v$ in $G$ through vertices anterior to $U'$ on which every interior vertex is a collider. Some vertex of $\rho$ is not anterior to $U$, since $uv$ is not an edge of $G_U$. Let $w$ be the first one. As $w$ is anterior to $U'$, it is the first vertex of a path that ends in $U'$ with edges undirected or directed towards that end. Let $b$ be the first vertex of $U'$ on this path, possibly $w$ itself. No vertex between $w$ and $b$ on this path is anterior to $U$, because $w$ is not.

The vertices $u$ and $w$ are adjacent in $G_{U\cup\{b\}}$, since the part of $\rho$ between them passes through vertices anterior to $U$. Consecutive vertices of the path from $w$ to $b$ are adjacent in $G$ and anterior to $b$, hence adjacent in $G_{U\cup\{b\}}$. Together they form the required path from $u$ to $b$.

(c) Suppose that some edge of $G_{U\cup\{l\}}$ joins a vertex anterior to $U$ to one that is not, and let $\rho$ be a corresponding path in $G$ between these two vertices. Let $w$ be the first vertex of $\rho$ that is not anterior to $U$ and $y$ the vertex before it. The edge between $y$ and $w$ is neither undirected nor directed from $w$ to $y$, as $w$ would be anterior to $U$. It therefore has an arrowhead at $w$, and $w$ has no undirected edge. Since $w$ is anterior to $U\cup\{l\}$ but not to $U$, it is anterior to $l$. Then $w=l$ or a directed path leads from $w$ to $l$. In both cases $l$ has an arrowhead, which an ancestral graph forbids for a vertex with an undirected edge.

(d) For $m=1,2$ let $\rho_m$ be a path from $x_m$ to $l$ in $G$ through vertices anterior to $U_m\cup\{l\}$ on which every interior vertex is a collider. Its last edge has an arrowhead at $l$. Indeed, it is not undirected, and if it were directed away from $l$ its other vertex would be a child of $l$ anterior to $U_m\cup\{l\}$. This child is not anterior to $U_m$, as $l$ would then be anterior to $U'$. So it is anterior to $l$ and has an arrowhead. A directed path from it to $l$ then closes a directed cycle.

Let $x_1\neq x_2$ and let $y$ be the first vertex of $\rho_1$ that lies on $\rho_2$. Follow $\rho_1$ from $x_1$ to $y$ and then $\rho_2$ from $y$ to $x_2$. This is a path through vertices anterior to $U'\cup\{l\}$. Each of its interior vertices other than $y$ is a collider on $\rho_1$ or on $\rho_2$. If $y$ is an interior vertex, its two edges on the new path have arrowheads at $y$, because $y$ equals $l$ or is a collider on the path concerned. Hence $x_1$ and $x_2$ are adjacent in $G_{U'\cup\{l\}}$.
\end{proof}

For a statement $t=\st{a}{b}{B}$ let $\Gamma_t$ be the graph obtained from $G_t+P_t$ by deleting the vertices of $B\setminus V(P_t)$. Then $t\in\cI(G)^P$ exactly when $a$ and $b$ lie in different connected components of $\Gamma_t$. If every index of $t$ occurs in a statement $u$, then $P_t\subseteq P_u$ and $G_t$ is a subgraph of $G_u$ by Lemma~\ref{lem:GU}(a).

\begin{lemma}\label{lem:several}
Let $G$ be an ancestral graph and $P$ a set of pairs.
\begin{enumerate}[label=\textup{(\alph*)},leftmargin=2.4em]
\item The set $\cI(G)^P$ is cancellation-free for every pair $\{k,l\}$.
\item Let $\{k,l\}\notin P$ and $P'=P\cup\{\{k,l\}\}$. A statement of $\cI(G)^P$ in which both $k$ and $l$ occur and that is screened in $\cI(G)^P$ belongs to $\cI(G)^{P'}$.
\end{enumerate}
\end{lemma}
\begin{proof}
(a) Let $i,j,k,l$ be distinct and $D\subseteq[n]\setminus\{i,j,k,l\}$. Suppose that $s_0=\st{i}{j}{D}$ and $s_2=\st{i}{j}{D\cup\{k,l\}}$ belong to $\cI(G)^P$. For $a\in\{i,j\}$ and $b\in\{k,l\}$ put $t_{ab}=\st{a}{b}{D}$. Assume that none of these four statements belongs to $\cI(G)^P$, so that $\Gamma_{t_{ab}}$ contains a path $\pi_{ab}$ from $a$ to $b$. We call $b\in\{k,l\}$ paired if it belongs to a pair of $P_{s_2}$. A contradiction is reached in four steps.

\emph{Step 1. The vertices $k$ and $l$ are not both paired.} Suppose they are. The vertices deleted in $\Gamma_{s_2}$ are then those of $D\setminus V(P_{s_2})$, and they are deleted in every $\Gamma_{t_{ab}}$. Hence $\Gamma_{t_{ab}}$ is a subgraph of $\Gamma_{s_2}$, in which $\pi_{ik}$ and $\pi_{jk}$ join $i$ to $j$, contrary to the assumption $s_2\in\cI(G)^P$.

\emph{Step 2. An unpaired vertex $b$ is not anterior to $D\cup\{i,j\}$.} The pairs of $P_{t_{ab}}$ avoid $b$, so they belong to $P_{s_0}$. If $b$ were anterior to $D\cup\{i,j\}$, the graph $G_{t_{ab}}$ would be a subgraph of $G_{s_0}$ by Lemma~\ref{lem:GU}(a), and $\Gamma_{t_{ab}}$ a subgraph of $\Gamma_{s_0}$. The paths $\pi_{ib}$ and $\pi_{jb}$ would join $i$ to $j$ in $\Gamma_{s_0}$, contrary to $s_0\in\cI(G)^P$.

\emph{Step 3. For an unpaired vertex $b$ some walk from $i$ to $j$ in $G_{s_2}+P_{s_2}$ avoids $b$ and $D\setminus V(P_{s_2})$.} The path $\pi_{ab}$ starts at a vertex anterior to $D\cup\{a\}$ and ends at $b$, which is not. The pairs of $P_{t_{ab}}$ lie inside $D\cup\{a\}$, so some edge of $\pi_{ab}$ is an edge of $G_{t_{ab}}$ that leaves the vertices anterior to $D\cup\{a\}$. By Lemma~\ref{lem:GU}(c) the vertex $b$ has no undirected edge. Let $x_a$ be the vertex before $b$ on $\pi_{ab}$, which is adjacent to $b$ in $G_{t_{ab}}$. By Lemma~\ref{lem:GU}(d) the vertices $x_i$ and $x_j$ are equal or adjacent in $G_{D\cup\{i,j,b\}}$, a subgraph of $G_{s_2}$. Joined by this edge, the part of $\pi_{ib}$ before $b$ and that of $\pi_{jb}$ form a walk $\omega_b$ from $i$ to $j$ in $G_{s_2}+P_{s_2}$. It avoids $b$ as well as the vertices of $D\setminus V(P_{s_2})$.

\emph{Step 4. Conclusion.} By Step 1 some $b\in\{k,l\}$ is unpaired. If the other vertex of $\{k,l\}$ is paired, $\omega_b$ is a walk in $\Gamma_{s_2}$, a contradiction. Otherwise $k$ and $l$ are both unpaired. By Steps 2 and 3 neither is anterior to $D\cup\{i,j\}$ or has an undirected edge. Since $G$ has no directed cycle, at most one of them is anterior to the other. Say $k$ is not anterior to $l$. Then $k$ lies in neither $G_{t_{il}}$ nor $G_{t_{jl}}$, so $\omega_l$ avoids $k$ and is a walk in $\Gamma_{s_2}$, again a contradiction.

(b) Let $s=\st{i}{j}{C}$ be screened by $e$, and put $D=C\setminus\{k,l\}$. We may assume $e=i$. Let $T=\{j,k,l\}\setminus\{i\}$ and $t_a=\st{i}{a}{D}$ for $a\in T$. The indices that occur in $s$ are those of $\{i\}\cup T\cup D$. The pairs of $P'$ inside $s$ are those of $P_s$ together with $\{k,l\}$, so the separating set of $s$ with respect to $P'$ is the set $D\setminus V(P_s)$.

Let $P_0$ be the set of pairs of $P$ inside $D\cup\{i\}$ and $\Gamma_0$ the graph obtained from $G_{D\cup\{i\}}+P_0$ by deleting $D\setminus V(P_0)$. For every $a\in T$ it is a subgraph of $\Gamma_{t_a}$, because $P_0\subseteq P_{t_a}$ and $G_{D\cup\{i\}}$ is a subgraph of $G_{t_a}$.

Suppose that $s\notin\cI(G)^{P'}$. Adding the edge $\{k,l\}$ to $G_s+P_s$ then gives a graph with a path $\pi$ from $i$ to $j$ that avoids $D\setminus V(P_s)$. Let $\pi_0$ be the longest initial part of $\pi$ that is a path in $\Gamma_0$ and contains no vertex of $T$. Let $u$ be its last vertex and $v$ the next vertex of $\pi$. We show that $i$ is joined to some $a\in T$ in $\Gamma_{t_a}$, which contradicts $t_a\in\cI(G)^P$.

\emph{Case 1: $uv$ is not an edge of $G_{D\cup\{i\}}+P_0$.} It is not the edge $\{k,l\}$, because $u\notin T$. If it is a pair of $P_s$, then $u\in D\cup\{i\}$ and $v$ is some $a\in T$. The pair lies inside $D\cup\{i,a\}$, and $\pi_0$ followed by it joins $i$ to $a$ in $\Gamma_{t_a}$. Otherwise $uv$ is an edge of $G_s$ that is not an edge of $G_{D\cup\{i\}}$. Lemma~\ref{lem:GU}(b) provides a vertex $a$ of $s$ that is not anterior to $D\cup\{i\}$, hence $a\in T$, and a path from $u$ to $a$ in $G_{t_a}$. Its vertices other than $u$ are not anterior to $D\cup\{i\}$, so they lie neither in $D$ nor on $\pi_0$. Following $\pi_0$ and then this path, we join $i$ to $a$ in the graph $\Gamma_{t_a}$.

\emph{Case 2: $uv$ is an edge of $G_{D\cup\{i\}}+P_0$.} By the choice of $\pi_0$, either $v\in T$ or $v\in D\setminus V(P_0)$. In the first case $\pi_0$ followed by $uv$ joins $i$ to $v$ in $\Gamma_{t_v}$. In the second case $v$ belongs to a pair of $P_s$, as $\pi$ avoids $D\setminus V(P_s)$. The other vertex of this pair is not in $D\cup\{i\}$, so it is some $a\in T$. The pair lies inside $D\cup\{i,a\}$, hence $v\in V(P_{t_a})$. Following $\pi_0$, the edge $uv$ and this pair, we join $i$ to $a$ in $\Gamma_{t_a}$.
\end{proof}

\begin{proof}[Proof of Theorem~\ref{thm:several}]
We first prove the claim for a Markov matrix $\Sigma$. Let $s=\st{i}{j}{C}\in\cI(G)^P$ and $D=C\setminus V(P_s)$. Let $A$ consist of the indices of $s$ outside $D$ that lie in the connected component of $i$ in $\Gamma_s$, and let $B$ consist of the other indices of $s$ outside $D$. Then $i\in A$ and $j\in B$. For $a\in A$ and $b\in B$ the set $D$ separates $a$ from $b$ in $G_s$, hence in its subgraph $G_{D\cup\{a,b\}}$, so that $\st{a}{b}{D}\in\cI(G)$ by the criterion recalled in Section~\ref{sec:several}. Hence $A\indep B\,|\,D$ holds for $\Sigma$ and $\Sigma_{A\cup D,B\cup D}$ has rank $|D|$.

A pair of $P$ with both vertices in $A\cup B\cup D$ belongs to $P_s$. Its vertices are adjacent in $\Gamma_s$, so both lie in $A$ or both lie in $B$. The submatrix therefore contains no changed entry, and $A\indep B\,|\,D$ holds for $\Sigma'$. Since $C=D\cup(A\setminus\{i\})\cup(B\setminus\{j\})$, the first property recalled in this appendix gives $s\in\ci{\Sigma'}$.

(a) Let $\Sigma$ be faithful to $G$. We use induction on the number of pairs. For $P=\emptyset$ the claim is the faithfulness of $\Sigma$, because $\cI(G)^{\emptyset}=\cI(G)$ by the same criterion. Let $P'=P\cup\{\{k,l\}\}$ with $\{k,l\}\notin P$, and let $\Sigma_1$ be obtained from $\Sigma$ by a change of the covariances of $P$ such that $\ci{\Sigma_1}=\cI(G)^P$. By Lemma~\ref{lem:several}(a) this set is cancellation-free for $\{k,l\}$. Proposition~\ref{thm:rule} then describes the structure of $\Sigma_1+\tau(E_{kl}+E_{lk})$ for generic admissible $\tau$. It consists of the statements of $\cI(G)^P$ that are untouched or screened. An untouched statement belongs to $\cI(G)^{P'}$, since $P$ and $P'$ have the same pairs inside it. A screened one does so by Lemma~\ref{lem:several}(b). With the first claim, the structure is $\cI(G)^{P'}$.

Consequently, for every statement $s\notin\cI(G)^{P'}$ the minor $\apm{s}$ of the matrix perturbed at the pairs of $P'$ is a non-zero polynomial in the sizes. Claim (a) for $P'$ thus holds for generic sizes.

(b) Apply (a) to the statements $\st{i}{j}{C}$ with $C=[n]\setminus\{i,j\}$, for which $P_s=P$ and $G_s=G^a$. The set $C\setminus V(P)$ fails to separate $i$ from $j$ in $G^a+P$ exactly when a path with interior in $V(P)$ joins them, that is, when they are adjacent in $(G^a+P)^{V(P)}$.
\end{proof}

\end{document}